\documentclass[twocolumn,trackchanges]{aastex63}

\newcommand{\inv}{$^{-1}$}
\newcommand{\Mpch}{$h^{-1}$ Mpc}

\usepackage{graphics}

\usepackage{amsmath}
\usepackage{amssymb}

\usepackage{mathrsfs}

\submitjournal{ApJ}

\shorttitle{Supermassive black hole fuelling in \texttt{IllustrisTNG}}
\shortauthors{Bhowmick et al.}

\begin{document}

\title{Supermassive black hole fueling in IllustrisTNG: Impact of environment}

\correspondingauthor{Aklant K. Bhowmick}
\email{aklantbhowmick@ufl.edu}

\author[0000-0002-7080-2864]{Aklant K. Bhowmick}
\affiliation{Department of Physics, University of Florida, Gainesville, FL 32611, USA} 

\author[0000-0002-2183-1087]{Laura Blecha}
\affiliation{Department of Physics, University of Florida, Gainesville, FL 32611, USA}

\author{July Thomas}
\affiliation{Department of Physics, University of Florida, Gainesville, FL 32611, USA}

\begin{abstract}

We study the association between active galactic nuclei (AGN) and 
environment at scales of $0.01-1$ \Mpch\ in the \texttt{IllustrisTNG} simulated universe (specifically, the TNG100 simulation).
We identify supermassive black hole (BH) pairs and multiples 
within scales of 0.01, 0.1, \& 1 \Mpch\ and examine their AGN activity 
in relation to randomly-selected pairs and multiples. The number density of BHs in TNG100 is 
$n = 0.06\,h^3$ Mpc$^{-3}$ at $z\lesssim1.5$ ($n = 0.02\, h^3$ Mpc$^{-3}$ at $z=3$). About $\sim10\%$ and $\sim1\%$ of these BHs live in pairs and multiples, respectively, within 0.1 \Mpch\ scales. We find that these systems have enhanced likelihood~(up to factors of 3-6) of containing high Eddington ratio~($\eta\gtrsim0.7$) AGN compared to random pairs and multiples. Conversely, the likelihood of an AGN to live in 0.1~\Mpch\ scale BH systems is also higher~(by factors $\sim4$ for $\eta\gtrsim0.7$) compared to random pairs and multiples. We also estimate that $\sim 10\%$ of ultra-hard X-ray selected AGN in TNG100 have detectable 2-10 keV AGN companions on 0.1~\Mpch\ scales, in agreement with observations. On larger spatial scales ($\sim 1$ \Mpch), however, no significant enhancements in AGN activity are associated with BH pairs and multiples, even at high Eddington ratios. The enhancement of AGN activity in 
rich, small-scale~($\lesssim0.1$ \Mpch) environments is therefore likely to be driven by 
galaxy interactions and mergers. 
Nonetheless, the overall percentage of AGN that live in $\lesssim0.1$ \Mpch\ scale multiples is still subdominant~(at 
most $\sim40\%$ for the highest Eddington ratio AGN). Furthermore, the associated enhancement in their Eddington ratios of BH systems~(as well as merging BHs) is only up to factors of $\sim2-3$. Thus, our results support the existence of a merger-AGN connection, but they also suggest that mergers and interactions play 
a relatively minor role in fueling the AGN population as a whole. 
\end{abstract}


\section{Introduction}
 
It is well established that supermassive black holes~(BHs) reside at the centers of almost all nearby massive galaxies~\citep{1992ApJ...393..559K,1994ApJ...435L..35H,1995Natur.373..127M}. We also know that a small fraction of galaxies have bright nuclei 
referred to as active galactic nuclei, or AGN, which are powered by accreting supermassive BHs. 
Determining the dominant mechanisms that drive AGN fueling and their connection to BH-galaxy co-evolution is an ongoing challenge.

Fueling BH accretion requires the availability of cold gas with low angular momentum. 
The large scale~($\gtrsim1$ \Mpch) and small scale~($\lesssim0.1$ \Mpch) environments of AGN can therefore provide important clues about their fueling mechanisms. 
For example, the weak dependence seen for the observed large scale clustering on AGN luminosity~\citep{2006MNRAS.373..457L,2018MNRAS.474.1773K,2019MNRAS.483.1452W,2020ApJ...891...41P} implies that more massive haloes do not necessarily host more luminous AGN, which is also seen in the large scatter in the AGN luminosity vs host halo mass relations in hydrodynamic simulations~\citep{2019MNRAS.485.2026B}. 
 Numerous mechanisms may contribute to AGN triggering. On one hand, this can be driven by secular processes occurring within the host galaxy such as supernova winds~\citep{2009ApJ...695L.130C,2010MNRAS.404.2170K} and hydrodynamic instabilities~\citep{2004ARA&A..42..603K,2011ApJ...741L..33B}. On the other hand, external disturbances to the host galaxy, such as tidal torques generated during galaxy interactions and mergers are also very promising candidates, particularly in gas-rich, major mergers \citep[e.g.,][]{1996ApJ...471..115B,2005Natur.433..604D,2008ApJS..175..356H, 2013MNRAS.429.2594B,2015MNRAS.447.2123C,2019SCPMA..62l9511Y}.



Currently, there is no clear consensus about whether galaxy mergers or secular processes are the dominant driver of BH fueling. 
A vast majority of AGN host galaxies do not exhibit any evidence of recent mergers~\citep{2014MNRAS.439.3342V,2019ApJ...882..141M}. Several works analysing the morphologies of the host galaxies found no significant differences in the `merger fractions' between active and inactive galaxies~\citep[e.g.,][]{2009ApJ...691..705G,2011ApJ...726...57C,2012ApJ...744..148K,2012MNRAS.425L..61S,2014MNRAS.439.3342V,2017MNRAS.466..812V,2019ApJ...882..141M,2019MNRAS.489..497Z,2019ApJ...877...52Z}. In contrast, many other works have also found that galaxies that do exhibit signatures of mergers or interactions have higher AGN fractions compared to those that do not~\citep[e.g.,][]{2011ApJ...737..101L,2011ApJ...743....2S,2011MNRAS.418.2043E,2013MNRAS.435.3627E,2014AJ....148..137L,2014MNRAS.441.1297S,2017MNRAS.464.3882W,2018PASJ...70S..37G,2019MNRAS.487.2491E}. 
Potential signatures of the AGN-merger connection can also be seen in small-scale quasar clustering measurements of binary quasar pairs, wherein enhanced clustering amplitude is reported at small scales~(few tens of kiloparsecs), as compared to extrapolations from large scale clustering~\citep{2000AJ....120.2183S,2006AJ....131....1H,2012MNRAS.424.1363K,2016AJ....151...61M,2017MNRAS.468...77E}. However, observational studies of merger-triggered AGN are associated with several challenges. 
For one, the AGN luminosity can make it difficult to identify morphological merger signatures in the host galaxy. 
Galaxy mergers may also create 
a significant amount of quasar obscuration, as seen in both simulations~\citep[e.g.,][]{2006ApJS..163....1H,2013ApJ...768..168S,2018MNRAS.478.3056B} and observations~\citep[e.g.,][]{1988ApJ...325...74S,1996ARA&A..34..749S,2009ApJS..182..628V,2017MNRAS.468.1273R}; this can make it difficult to identify merger-triggered AGN. The resulting systematic biases could potentially explain the seemingly conflicting results in the existing literature; this is corroborated by the growing evidence of high merger fractions amongst obscured AGN~\citep[e.g.,][]{2008ApJ...674...80U,2015ApJ...806..218G, 2015ApJ...814..104K,2018Natur.563..214K,2020arXiv200400680G}.      
Cosmological hydrodynamic simulations have also shown statistically robust evidence of the presence of the merger-AGN connection. A recent study by \cite{2020MNRAS.494.5713M} looked at merger fractions of AGN hosts as well as AGN fractions of merging galaxies within the EAGLE simulation~\citep{2015MNRAS.446..521S} and demonstrated the existence of a merger-AGN connection, though they also found that merger driven activity does not contribute significantly to the overall growth history of the BH populations. Using data from the MassiveBlackII simulation, \citep{2019MNRAS.485.2026B, 2020MNRAS.492.5620B} demonstrated that merger-driven AGN activity also leads to the formation of systems of multiple active AGN. 
Additionally, in a companion paper to the present work, Thomas et al. (in prep) are using very high time resolution data from IllustrisTNG to quantify merger-driven AGN fueling in detail.

An inevitable consequence of hierarchical clustering of halos (and the eventual merging of their member 
galaxies) is the formation of systems of multiple BHs. These processes involve several stages which together encompass a huge dynamic range~($\sim9$ orders of magnitude) of separation scales between the BHs. The earliest stages are marked by the gravitational clustering of dark matter halos involving scales $\sim1-100~\mathrm{Mpc}$. The next stage can be marked by when the halos merge and their respective galaxies start interacting; this occurs at scales of $\sim 100$ kpc. The galaxies then eventually merge via dynamical friction~\citep{1943ApJ....97..255C}. 
Following the galaxy merger, dynamical friction causes the BHs to continue to inspiral until they reach parsec scales. 
The timescales for further hardening of the ensuing BH binaries to scales below $\sim1$ pc are uncertain and may be many Gyr in some cases~\citep{1980Natur.287..307B,2003AIPC..686..201M}; this is known as the ``Final Parsec problem."
Binaries may evolve on these scales via repeated three body scatterings with stars~\citep{1996NewA....1...35Q,2015ApJ...810...49V,2017MNRAS.464.2301G,2020MNRAS.493.3676O} as well as via interactions with gas~\citep{2002ApJ...567L...9A,2014ApJ...794..167S,2016ApJ...827..111R}. If the binaries reach sufficiently small scales~($\sim$ a few mpc), gravitational wave (GW) radiation will take over and cause the BHs to merge; these GWs may be detectable with current and upcoming facilities such as pulsar timing arrays (PTAs) 
\citep[e.g.,][]{2013PASA...30...17M,2016MNRAS.458.3341D,2016MNRAS.458.1267V,2019BAAS...51g.195R} as well as the Laser Interferometer 
Space Antenna (LISA) \citep{2019arXiv190706482B}. Cosmological hydrodynamic simulations enable us to probe the formation and evolution of such BH systems from $\sim1~\mathrm{Mpc}$ to $\sim0.01~\mathrm{Mpc}$ scales (the 
resolution of the simulation prevents us from probing scales smaller than 
$\sim 0.01~\mathrm{Mpc}$). These correspond to relatively early stages of galaxy mergers, which are precursors to gravitational bound BH binaries~(BHBs) that will be powerful GW sources for 
LISA and PTAs. 
Numerous recent models based on simulations or semi-analytic modeling have made detailed predictions about the 
formation and evolution of BHBs~\citep{2010ApJ...719..851S,2013ApJ...773..100K,2014MNRAS.442...56R,2015ApJ...810..139H,2016MNRAS.461.4419B,2017MNRAS.464.3131K,2019MNRAS.486.4044B,2019ApJ...887...35M,2020arXiv200414399N}. However, connecting these models to observations continues to be a challenge. Current statistical samples of close BH pairs are largely between $\sim 1-100~\mathrm{kpc}$ scale separations  \citep[e.g.,][]{2011MNRAS.418.2043E,2011ApJ...737..101L,2011ApJ...735L..42K,2013ApJ...777...64C,2019ApJ...882...41H,2019ApJ...883..167P,2020arXiv200110686H}. In contrast, only one confirmed parsec-scale BH \textit{binary} is known~\citep{2006ApJ...646...49R}, and the growing population of unresolved, mpc-scale binary candidates requires extensive follow-up for confirmation~\citep{2019ApJ...884...36L,2020MNRAS.494.4069K}.
Therefore, while these early-stage, $\sim 1-100~\mathrm{kpc}$ scale BH pairs are still at separations much larger than the GW regime, their properties can serve as an important baseline for BHB models to make predictions on the overall abundances of BHBs and their electromagnetic signatures.



In this work, we use the TNG100 simulation from the \texttt{Illustris-TNG} simulation suite to investigate the possible association between AGN activity 
and the richness of the AGN environment 
at a wide range of scales~($0.01-1$ \Mpch). For our purposes, we measure ``environmental richness" in terms of BH multiplicity, or the abundance of nearby BHs. In the process, we explore the possibility of enhanced AGN activity associated with multiple BH systems, which~(if it exists) may be attributed to a range of physics including 
1) large-scale ($\gtrsim$ 1 \Mpch) clustering of massive haloes hosting luminous AGN 
and 2) galaxy mergers and interactions on $\lesssim$ 0.1 \Mpch\ scales producing luminous AGN. 
In particular, we identify systems of multiple BHs within separations of 0.01, 0.1, \& 1.0 \Mpch\ and investigate AGN activity of these multiples as compared to isolated BHs. We also examine AGN activity in merging BHs, based on the recorded time of BH merger rather than the final pre-merger BH pair separation resolved in the simulation. We do not classify multiple BH systems based on host galaxy properties such as stellar mass ratio \citep[in contrast to the recent study of][]{2020MNRAS.494.5713M}; instead, we focus solely on AGN activity as a function of relative BH positions and merger times. In addition to its simplicity, our approach avoids the uncertainty in measuring stellar masses of close or interacting systems, as tidal stripping tends to strongly alter the mass ratio between first infall and merger~\citep{2016MNRAS.458.2371R,2017MNRAS.464.1659Q}.
In addition, 
our analysis of multiple BH systems on 0.01 - 1 Mpc scales is complementary to the approach in
our companion paper (Thomas et. al.~ in prep.), which provides an in-depth analysis of merger-triggered BH growth 
using higher time resolution BH data. 
In Section \ref{methods}, we describe our basic methodology which includes a brief description of \texttt{Illustris-TNG}, as well as the criteria used for the identification of BH systems. Section \ref{bh_systems} presents some basic properties of the BH systems, particularly the relationship with their host halos, as well as their abundances. Section \ref{AGN_activity} focuses on the AGN activity of these BH systems. Section \ref{conclusions} summarizes the main results and conclusions.    


%

\section{Methods}
\label{methods}
\subsection{\texttt{Illustris-TNG} simulation}
The \texttt{Illustris-TNG} project~\citep[e.g.,][]{2018MNRAS.473.4077P,2018MNRAS.475..624N,2018MNRAS.480.5113M,2019MNRAS.490.3196P,2019MNRAS.490.3234N} is a suite of large 
cosmological magnetohydrodynamics~(MHD) simulations with three cosmological volumes: \texttt{TNG50}, \texttt{TNG100} and \texttt{TNG300}, corresponding to box lengths 
of $50$, $100$ and $300$ comoving Mpc, respectively. The \texttt{Illustris-TNG} simulations are successors to the original \texttt{Illustris} simulation~\citep[e.g.,][]{2014MNRAS.444.1518V,2015A&C....13...12N},  with improved subgrid physics modeling that produces more realistic galaxy populations in better agreement with observations 
\citep[e.g.,][]{2018MNRAS.479.4056W,2018MNRAS.475..648P,2018MNRAS.475..676S,2019arXiv190407238V}.
The simulation was run using the moving mesh code \texttt{AREPO}~\citep{2010MNRAS.401..791S,2011MNRAS.418.1392P,2016MNRAS.462.2603P}, which solves for self-gravity coupled with MHD. 
The gravity solver uses the PM-tree method~\citep{1986Natur.324..446B} whereas the fluid dynamics solver uses a finite volume Godunov scheme in which
the spatial discretization is performed using an unstructured, moving Voronoi tessellation of the domain. The base cosmology is adopted from the results of \cite{2016A&A...594A..13P} which is summarized by the following set of parameters: $\Omega_\Lambda=0.6911,~\Omega_m=0.3089,~\Omega_b=0.0486,~H_0=67.74~\mathrm{km~sec^{-1}~Mpc^{-1}},~\sigma_8=0.8159,~n_s=0.9667$. These cosmological parameters are assumed throughout this work. The simulations were initialised at $z=127$ using glass initial conditions~\citep{1994astro.ph.10043W} 
along with the Z'eldovich approximation~\citep{1970A&A.....5...84Z} to construct the initial displacement field.  

In addition to the gravity and MHD, 
the simulation includes a wide array of physics to model the key processes responsible for galaxy formation and evolution. Due to resolution limitations, the implementation is carried out in the form of `sub-grid' recipes that include the following:
\begin{itemize}
\item Star formation in a multiphase interstellar medium~(ISM) based on the prescription in \cite{2003MNRAS.339..289S}, with inclusion of chemical enrichment and feedback from supernovae (SNe) and stellar winds as described in \cite{10.1093/mnras/stx2656}.

\item Cooling of metal-enriched gas in the presence of a redshift dependent, spatially uniform, ionizing UV background, with self-shielding in dense gas as described in \cite{2013MNRAS.436.3031V}.

\item Magnetic fields are included via a small uniform initial seed field~($\sim10^{-14}$ Gauss) at an arbitrary orientation~\citep{2018MNRAS.480.5113M}. The subsequent evolution~(coupled with the gas) is driven by the equations of magnetohydrodynamics.

\item BH growth via gas accretion and mergers, as well as AGN feedback, which we describe in the following section. 
\end{itemize}

Halos are identified using a Friends-of-Friends~(FOF) algorithm~\citep{1985ApJ...292..371D} with a linking length equal to 0.2 times the mean particle separation. Within these halos, self-bound substructures~(subhalos) are identified using \texttt{SUBFIND}~\citep{2001NewA....6...79S}. 

\subsection{BH growth and AGN feedback in \texttt{Illustris-TNG}}
\begin{table*}
    \centering
    \begin{tabular}{|c|c|c|c|c|c|c|c|}
    \hline
$z$ & $n^{\mathrm{bh}}$&  $d_{\mathrm{max}}$ & $f^{\mathrm{sys}}(\mathscr{M}=1)$ & $f^{\mathrm{sys}}(\mathscr{M}=2)$ & $f^{\mathrm{sys}}(\mathscr{M}=3)$ & $f^{\mathrm{sys}}(\mathscr{M}=4)$& $f^{\mathrm{sys}}(\mathscr{M}>4)$ \\

& [$h^3$ Mpc$^{-3}$] & [$h^{-1}$ Mpc] & & & & &\\
\hline

0  & 6.01e-02 & 1.0 & 25.6 $\%$ & 13.0 $\%$ & 7.77 $\%$ & 5.33 $\%$ & 48.3 $\%$ \\

\texttt{"} & \texttt{"} & 0.1  & 88.8 $\%$ & 8.83 $\%$ & 1.75 $\%$ & 0.379 $\%$ & 0.185 $\%$ \\

\texttt{"} & \texttt{"} & 0.01  & 99.5 $\%$ & 0.458 $\%$ &  &  &  \\

\hline
0.6 & 5.74e-02 & 1.0  & 25.5 $\%$ & 13.9 $\%$ & 8.31 $\%$ & 6.79 $\%$ & 45.5 $\%$ \\

\texttt{"} & \texttt{"} & 0.1  & 89.3 $\%$ & 9.00 $\%$ & 1.38 $\%$ & 0.314 $\%$ & 0.0410 $\%$ \\

\texttt{"} & \texttt{"} & 0.01  & 99.7 $\%$ & 0.306 $\%$ &  &  & \\

\hline
1.5 & 4.65e-02 & 1.0  & 29.0 $\%$ & 16.0$\%$ & 9.84 $\%$ & 6.60 $\%$ & 38.5 $\%$ \\

\texttt{"} & \texttt{"}  & 0.1& 89.6 $\%$ & 8.78 $\%$ & 1.16 $\%$ & 0.387 $\%$ & 0.0810 $\%$ \\

\texttt{"} & \texttt{"} & 0.01  & 99.7 $\%$ & 0.285 $\%$ &  & &\\

\hline
3 & 1.95e-02 & 1.0  & 41.5 $\%$ & 18.1$\%$ & 10.7 $\%$ & 7.00 $\%$ & 22.7 $\%$ \\

\texttt{"} & \texttt{"} & 0.1 & 91.1 $\%$ & 7.88 $\%$ & 0.875 $\%$ & 0.146 $\%$ &  \\

\texttt{"} & \texttt{"} & 0.01  & 99.7 $\%$ & 0.267 $\%$ & & &\\
\hline
\hline

    \end{tabular}
    \caption{The overall abundances and BH systems in TNG100 in terms of number densities~(in units of $h^3~\mathrm{Mpc}^{-3}$). $n^{\mathrm{bh}}$~(3rd column) is the  number density of BHs. $f^{\mathrm{sys}}$~(4th-8th columns) is the percentage of BH singles~($\mathscr{M}=1$), pairs~($\mathscr{M}=2$), triples~($\mathscr{M}=3$), quadruples~($\mathscr{M}=4$) and beyond~($\mathscr{M}>4$) at various redshift snapshots~($z$) and separation scales~($d_{\mathrm{max}}$, in comoving \Mpch) within the \texttt{TNG100} simulation box.  The percentages~(in parentheses) refer to the fraction of BHs living as BH singles, pairs and multiples.}
    \label{number_of_systems}
\end{table*}
BHs of mass $8\times10^5~M_{\odot}~h$\inv\ are seeded in halos of total mass $>5\times10^{10}~M_{\odot}~h$\inv\ that do not already contain a BH. Once seeded, these BHs grow via Eddington-limited Bondi-Hoyle accretion given by
\begin{eqnarray}
\dot{M}_{\mathrm{BH}}=\mathrm{min}(\dot{M}_{\mathrm{Bondi}}, \dot{M}_{\mathrm{Edd}})\\
\dot{M}_{\mathrm{Bondi}}=\frac{4 \pi G^2 M_{\mathrm{BH}}^2 \rho}{c_s^3}\\
\dot{M}_{\mathrm{Edd}}=\frac{4\pi G M_{\mathrm{BH}} m_p}{\epsilon_r \sigma_T}c
\end{eqnarray}
where $G$ is the gravitational constant, $M_{\mathrm{BH}}$ is the mass of the BH, $\rho$ is the local gas density, $c_s$ is the local sound speed of the gas, $m_p$ is the mass of the proton, $\epsilon_r$ is the radiative efficiency and $\sigma_T$ is the Thompson scattering cross section. Accreting BHs radiate with a bolometric luminosity given by 
\begin{equation}
    L=\epsilon_r \dot{M}_{\mathrm{BH}} c^2,
\end{equation}
 with an assumed radiative efficiency of $\epsilon_r=0.2$.

A fraction of the energy released gets coupled to the surrounding gas as thermal or kinetic feedback. \texttt{Illustris-TNG} implements a two-mode feedback model as described in \cite{2017MNRAS.465.3291W}, the key features of which are summarized as follows. If the Eddington ratio~(defined as $\eta\equiv \dot{M}_{\mathrm{BH}}/\dot{M}_{\mathrm{edd}}$) exceeds a critical value of  $\eta_{\mathrm{crit}}=\mathrm{min}[0.002(M_{\mathrm{BH}}/10^8 M_{\odot})^2,0.1]$, thermal energy is injected into the neighboring gas at a rate given by $\epsilon_{f,\mathrm{high}} \epsilon_r \dot{M}_{\mathrm{BH}}c^2$, with $ ~\epsilon_{f,\mathrm{high}} \epsilon_r=0.02$. $\epsilon_{f,\mathrm{high}}$ is referred to as the ``high accretion state" coupling efficiency. If the Eddington ratio is below this critical value, kinetic energy is injected into the gas at regular intervals of time, in the form of a `wind' oriented along a randomly chosen direction. The injected energy is given by $\epsilon_{f,\mathrm{low}}\dot{M}_{\mathrm{BH}}c^2$ where $\epsilon_{f,\mathrm{low}}$ is referred to as the `low accretion state' coupling efficiency. $\epsilon_{f,\mathrm{low}}$ is assigned to have a maximum value of 0.2 with smaller values at very low gas densities. For further details on both feedback modes, we encourage the interested reader to refer to \cite{2017MNRAS.465.3291W}. 

An accurate modelling of BH dynamics at small scales is difficult because of the finite simulation resolution, which can result 
in spurious accelerations imparted to the BHs by numerical noise. To avoid this, BHs are (re)-positioned to the local potential minimum within a sphere containing $n$ 
neighboring gas cells, where $n=1000$ is the value assigned for \texttt{Illustris-TNG}. Such a repositioning naturally leads to a prompt merging of two BHs shortly after their parent subhalos merge. As discussed in detail below, this limits our ability to study unmerged BH systems on 0.01 \Mpch\ scales and motivates our inclusion of merging BH systems in parts of our analysis. We therefore avoid drawing statistical conclusions about these smallest-scale BH pairs and multiples.

\subsection{Identifying systems of BHs}
We identify BH systems by linking individual BHs within a maximum distance scale denoted by $d_{\mathrm{max}}$. In particular, every member 
of a BH system must be within a comoving distance $d_{\mathrm{max}}$ with respect to at least one other member.

We investigate systems at three values of $d_{\mathrm{max}}$:  1.0, 0.1, \& 0.01 \Mpch. $d_{\mathrm{max}}=1$ \Mpch\ roughly corresponds to typical distances between BHs in halos that have come together via gravitational clustering and are close to a merger; typically, the occupying galaxies themselves are not yet close enough to be visibly interacting. $d_{\mathrm{max}}=0.1$ \Mpch\ roughly corresponds to typical distances in the early stages of galaxy interactions, while
$d_{\mathrm{max}}=0.01$ \Mpch\ corresponds to typical distances in late-stage galaxy mergers.
It is important to also note that at the smallest $<0.01$ \Mpch\ scales, our samples are highly incomplete, because a significant portion of pairs at these scales are promptly merged by the BH repositioning scheme. As discussed below, we find that we are nonetheless able to draw useful qualitative conclusions from the population of unmerged small scale BH pairs. We additionally perform a more quantitative analysis of BHs in late-stage galaxy mergers using the more complete sample of BH merger progenitors (which are defined based on merger time, not BH separation). We shall be comparing the results for $<0.01$ \Mpch\ BH systems to those for 
the merging BHs, thereby 
allowing us to identify any systematic bias that may 
exist within the $<0.01$ \Mpch\ multiple-BH systems due to their 
incompleteness.

We define \textit{multiplicity}~(denoted by $\mathscr{M}$) as the number of members within a BH system. We characterize the AGN activity of a BH system by the member having the highest Eddington ratio; we shall refer to this as the \textit{primary} member of the system. Note that traditionally, the primary is defined to be the most massive BH. However, a merger triggered enhancement in the AGN activity does not necessarily occur within the most massive member. Therefore, our choice of the highest Eddington ratio member as the primary ensures that within every BH system, we are probing the BHs that are most likely to be associated with merger-triggered AGN (independent of BH mass). We also note that~(unless otherwise stated), our results are qualitatively independent of the choice of the primary~(the only exception being in Section \ref{impact_on_edd} and is discussed there).  


In order to study BH systems on both large and small spatial scales, we need a combination of high enough resolution as well as large enough volume to include a population of rare BH multiples. 
Therefore, in this work, we use 
the highest resolution realization of the \texttt{TNG100} box with $2\times1820^3$ 
resolution elements (DM particles and gas cells).

\subsection{Constructing randomized samples of BH systems}
\label{randomized_samples}
In order to analyse possible sources of selection bias in our results, we prepare an ensemble of ``randomized samples" of BH systems. Each randomized sample is constructed by randomly shuffling the ``system IDs"~(a unique integer ID we assign to each BH that determines which BH system it belongs to, if any) 
amongst all the BHs in the simulation. In other words, each system ID is assigned to a random BH within the simulation box. Therefore, for every BH system with multiplicity $\mathscr{M}$, there exists a subset of $\mathscr{M}$ randomly assigned BHs within each randomized sample. 
Using this procedure, for every sample of BH systems, we constructed an ensemble of 10 corresponding randomized samples that 
have identical abundances by construction for all multiples~(pairs, triples, and beyond). 

The following terminology is used for the remainder of the paper. We shall often refer to the actual BH systems within 1, 0.1, \& 0.01 \Mpch\ scales as ``true systems", and thereby compare their properties to their corresponding ``randomized systems" / ``randomly selected sets of BHs". Any selection bias in the computation of a quantity for a true system will be fully captured in the trends exhibited by the randomized systems.

\label{bh_systems}

\section{BH systems in \texttt{TNG100}}
\begin{figure*}
\centering
\includegraphics[width=18cm]{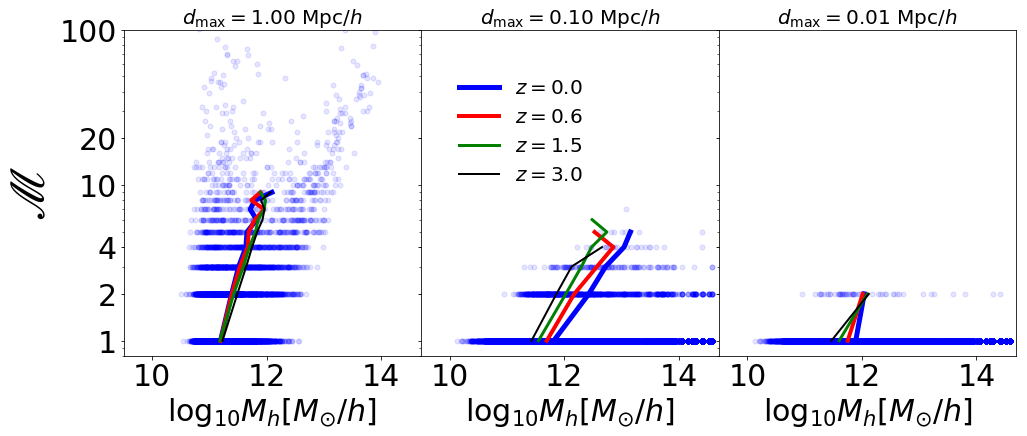}
\caption{Multiplicity~($\mathscr{M}$) as a function of the host halo mass of the primary BH~(member with the highest Eddington ratio), for BH systems~(all members have masses $>10^6~h^{-1}~M_{\odot}$) as predicted by the TNG100 simulation.~The blue circles correspond to the scatter at $z=0$. The colored solid lines show the mean trends at $z=0,1.5,3,4$. The different panels correspond to different values of $d_{\mathrm{max}}$, which is the maximum comoving distance between a member 
and at least one other member within the system. We see that more massive halos tend to host richer systems of BHs.}
\label{richness_vs_halomass}
\end{figure*}


Table \ref{number_of_systems} summarizes 
the abundances of the BH systems within the \texttt{TNG100} box. Here we discuss some of the basic properties~(environment and number densities) of these BH systems. Figure \ref{richness_vs_halomass} shows the multiplicity vs. host halo mass of the BH systems identified within the \texttt{TNG100} universe. Note that for $d_{\mathrm{max}}=1$ \Mpch, not all members will necessarily be within the same halo; in this case we choose the host halo mass of the primary member. Across all scales, we find that systems with higher multiplicity tend to live in more massive halos. This is simply a consequence of higher BH occupations in more massive halos. At scales within $1$ \Mpch~(leftmost panel), BH pairs, triples, and quadruples primarily reside in halos with total mass 
between $10^{11}-10^{13}~ h^{-1} M_{\odot}$, whereas the more massive $M_h\gtrsim10^{13}~h^{-1} M_{\odot}$ halos tend to host BH systems with  $\mathscr{M} \gtrsim10$. The median halo masses ($M_h\sim10^{11.5-12}~h^{-1} M_{\odot}$) of 
BH systems on these scales vary little with multiplicity. 
On smaller scales ($\leq0.1$ \Mpch, middle panel), a stronger trend with $M_h$ is seen; 
the median halo mass for pairs and triples is $\sim 10^{12}~h^{-1} M_{\odot}$, while the highest order multiples at these scales contain up to $\sim5$ members and have median halo masses of $\sim10^{13}~h^{-1} M_{\odot}.$ Finally, at scales of 
$\leq0.01$ \Mpch~(rightmost panel), no higher-order ($\mathscr M>2$) systems are present; BH pairs have 
median halo masses of $M_h\sim10^{12}~h^{-1} M_{\odot}$. 

\begin{figure*}
\centering
\includegraphics[width=18cm]{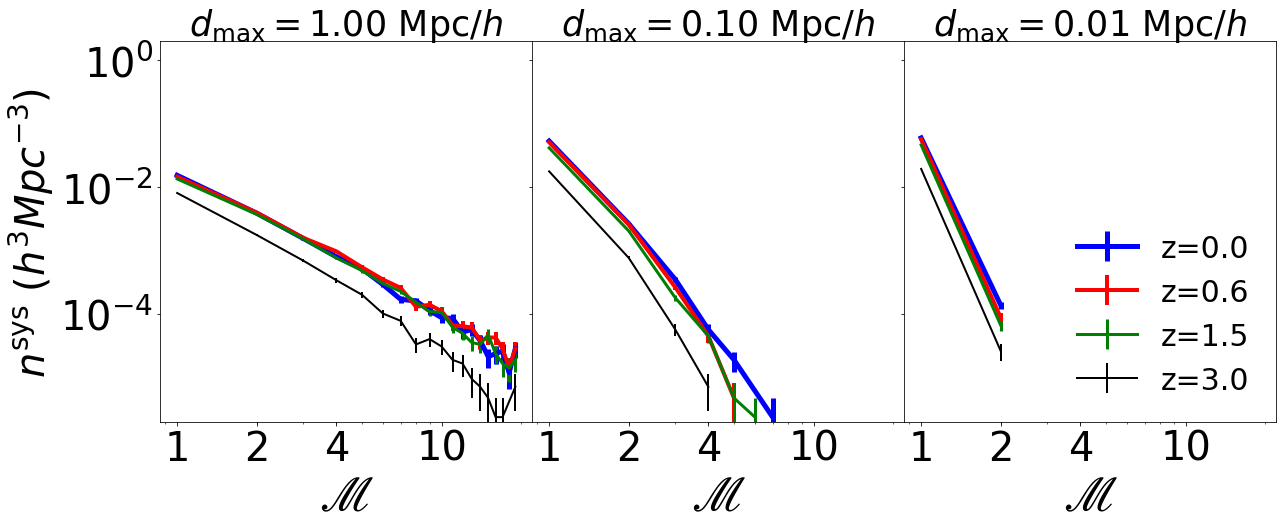}

\caption{Solid lines show the BH multiplicity functions~(defined as the number density $n^{\mathrm{sys}}$ of BH systems at each 
multiplicity $\mathscr{M}$), 
as predicted by the TNG100 simulation. Horizontal dotted lines mark the overall number density of BHs. Different colors correspond to snapshots at different redshift 
between $z=0-3$. Left to right, panels correspond to $d_{\mathrm{max}}=1$, 0.1, \& 0.01 \Mpch. The error bars correspond to Poisson errors. The number density of multiple BH systems decreases~(roughly as a power law) with increasing multiplicity, and higher-$\mathscr{M}$ BH systems are increasingly rare on small scales (lower $d_{\mathrm{max}}$). Additionally, we also construct 10 randomized samples of BH systems~(see Section \ref{methods} for how they are constructed) wherein the BHs are randomly grouped together to form systems such that they have the same multiplicity functions as the actual sample.}
\label{bh_multiplicity_functions}
\end{figure*}

\begin{figure*}

\includegraphics[width=18cm]{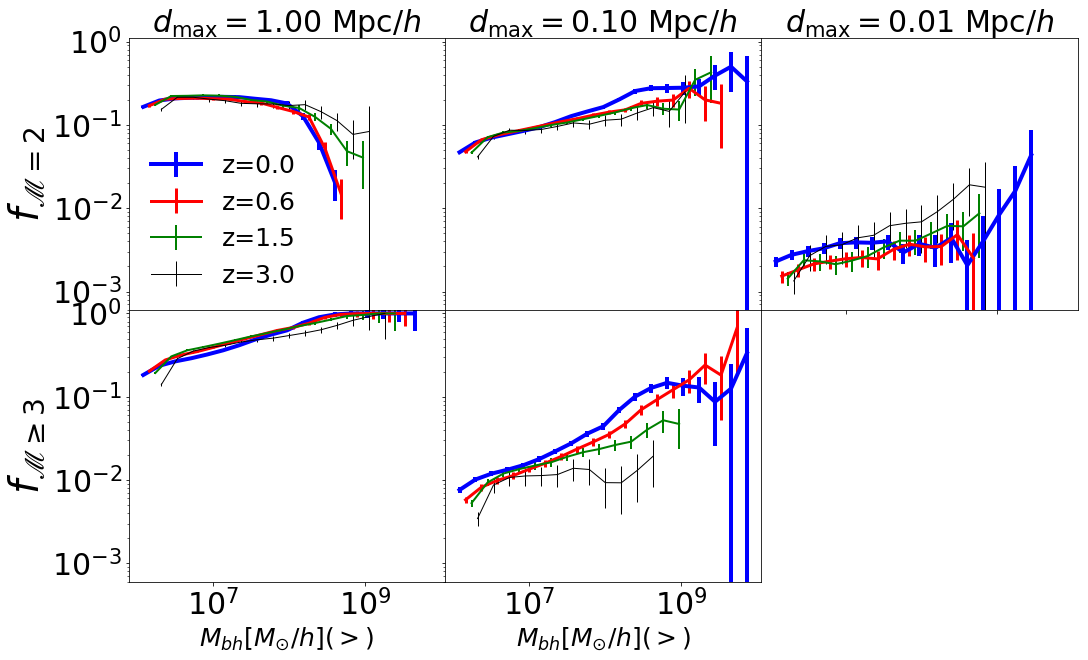}

\caption{\textbf{Upper Panels:} $f_{\mathscr{M}=2}$ is the fraction of systems that are pairs, plotted as a function of BH mass threshold~(defined by the mass of the most massive member).  Left to right, panels correspond to $d_{\mathrm{max}}=1.0$, 0.1, \& 0.01 \Mpch, respectively. Different colors correspond to snapshots at different redshifts 
between $z=0$ and 3. \textbf{Lower Panels:} Similar to the top panels, but $f_{\mathscr{M}\geq3}$ is the fraction of systems that are triples and higher order multiples. We find that more massive BHs have a higher likelihood of being members of multiple BH systems.}
\label{fraction_most_masive}

\end{figure*}

Now we consider the redshift evolution of the multiplicity versus halo mass relation from $z\sim0-3$. We focus on scales within 0.1 and $1$ \Mpch, where there are sufficient statistics for analysis. For BH systems within $0.1$ \Mpch\ scales, BH multiples at fixed halo mass are somewhat more common at higher redshift. 
In part, this reflects the fact that many such systems merge between $z=3$ and $z=0$. For $1$ \Mpch\ scales, no significant redshift evolution is seen, likely because more of these large-scale BH multiples are still unmerged at $z=0$.


Figure \ref{bh_multiplicity_functions} shows the volume density of the BH systems as a function of multiplicity. At $z=0$, 
the number densities for simulated BH pairs~($\mathscr{M}=2$) are 
$3.9\times10^{-3}$, $2.7\times10^{-3}$ and $1.4\times10^{-4}~h^3\mathrm{Mpc}^{-3}$ at scales of 1.0, 0.1, \& 0.01 \Mpch, respectively. Comparing this to the number density of the overall BH population at $z=0$ ($6 \times 10^{-2}~h^3\mathrm{Mpc}^{-3}$), we see that $\sim13\%$ of BHs live in pairs within 1 \Mpch\ scales. At smaller scales ($\leq0.1$ \& 0.01 \Mpch), the percentages of BHs in pairs decrease to $\sim9\%$ and $\sim0.5\%$, respectively. 

Amongst the pairs, $\sim37\%$ and $\sim17\%$ have additional companions to form triples within scales of $1$ \Mpch\ and $0.1$ \Mpch, respectively. Some of these systems may 
eventually form gravitationally bound triple BH systems, which may induce rapid BH mergers and provide a possible solution to the so-called ``final parsec problem"~\citep{2016MNRAS.461.4419B,2018MNRAS.473.3410R}. In addition to offering exciting prospects for gravitational wave detections, strong triple BH interactions will often eject the lightest BH from the system, creating a possible population of wandering BHs in galaxy halos \citep{Perets_2008,2010ApJ...721L.148B}.  

For higher order multiples, we see an approximate power-law decrease in the abundance of BH systems with increasing multiplicity for all values of $d_{\mathrm{max}}$. At smaller $d_{\mathrm{max}}$, there are fewer (or no) systems of multiple BHs, which leads to an increasingly sharp decline 
as we go from $d_{\mathrm{max}}=1$ \Mpch\ to $0.01$ \Mpch. The number densities of BHs and corresponding BH systems are nearly constant at $z<1.5$, while at $z=3$ they are lower by a factor of $\sim 3$. For a given value of $d_{\mathrm{max}}$, the relative proportion of BH singles and pairs is nearly constant from $z=0-3$, and the number density of $\mathscr{M}>2$ systems at $z=0$ is only slightly higher~(by $\sim10\%$ \& $60\%$ at $d_{\mathrm{max}}=0.1~\&~0.01$ \Mpch, respectively) than at $z>0.6$.

We now focus on how BH multiplicity 
depends on BH mass. We do this by looking at the relationship between the multiplicity and the most massive member of the system;  this is shown in Figure \ref{fraction_most_masive} for pairs~($\mathscr{M}=2$) and multiples~($\mathscr{M}\geq3$). Let us first focus on systems that are exclusively pairs~($\mathscr{M}=2$). At scales within $0.1$ \Mpch, the percentage of pairs increases with BH mass from $\sim5\%$ for $\sim10^6~h^{-1}~M_{\odot}$ BHs to $\sim20-40\%$ for $\sim10^9~h^{-1} M_{\odot}$ BHs. At scales within $0.01$ \Mpch, $\sim0.2-2\%$ of BHs live in pairs across the entire range of BH masses; there is some hint of increase in multiplicity with BH mass, although the statistics are very limited. At scales within $1$ \Mpch, we see that the percentage of pairs remains largely constant at $\sim20\%$ up to $\sim10^8~h^{-1}~M_{\odot}$, and then drops down to $\lesssim10\%$ at $\sim10^9~h^{-1}~M_{\odot}$; this is because at these scales, as we increase the mass of BHs, they have a much higher tendency of living in multiples~($\mathscr{M}\geq3$) instead of pairs. 


At scales within $0.1$ \Mpch, the percentage of BHs living in multiples with $\mathscr{M}\geq3$ is $\sim1-2\%$ for  $\sim10^7~h^{-1}~M_{\odot}$ BHs; this increases up to  
$\sim20\%$ for 
$\sim10^9~h^{-1}~M_{\odot}$ BHs. 
At $\leq1$ \Mpch\ scales, the percentage increases from $\sim30\%$ for $\sim10^7~h^{-1}~M_{\odot}$ BHs to $\sim60\% $ for $\sim10^8~h^{-1}~M_{\odot}$ BHs; almost all BHs with $\sim10^9~M_{\odot}$ and higher live within $1$ \Mpch\ scale pairs. 
Overall, we find that higher-mass BHs are more likely to have companions, as they live in more massive halos (cf.~Figure \ref{richness_vs_halomass}). Note that this also means that higher-multiplicity systems will have higher bolometric luminosities, on average. 
This motivates our choice~(in the following sections) to characterize the luminosity of the BHs in terms of their Eddington ratios; the Eddington ratios do not correlate as strongly with BH mass, making them a better proxy for the probing the AGN activity~(independent of the trends seen with BH mass).

\section{AGN activity within BH systems}
\label{AGN_activity}
\begin{figure*}
\centering
\includegraphics[width=13.5cm]{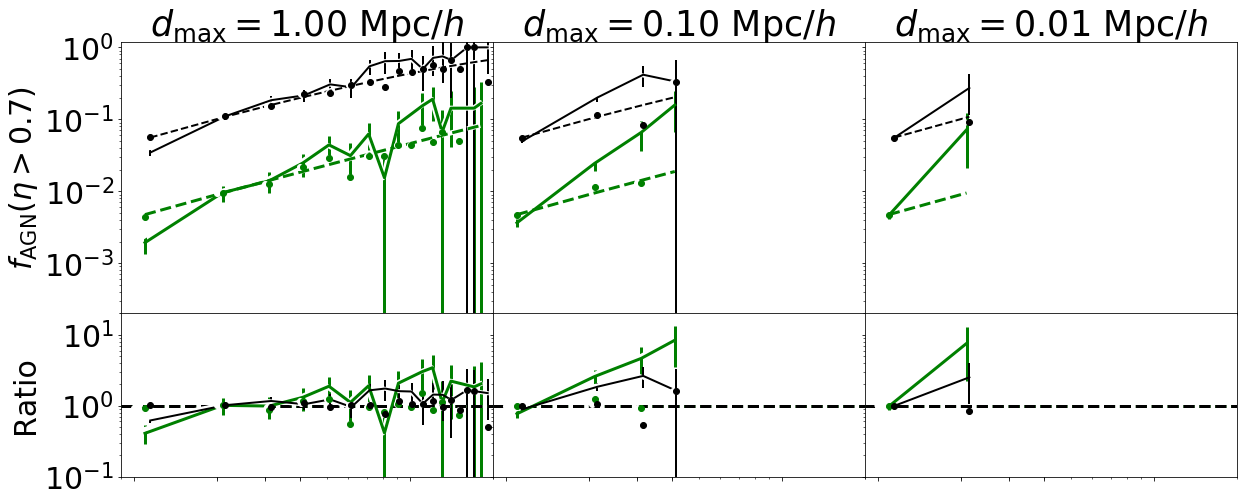}
\vspace{0.5 cm}

\includegraphics[width=13.5cm]{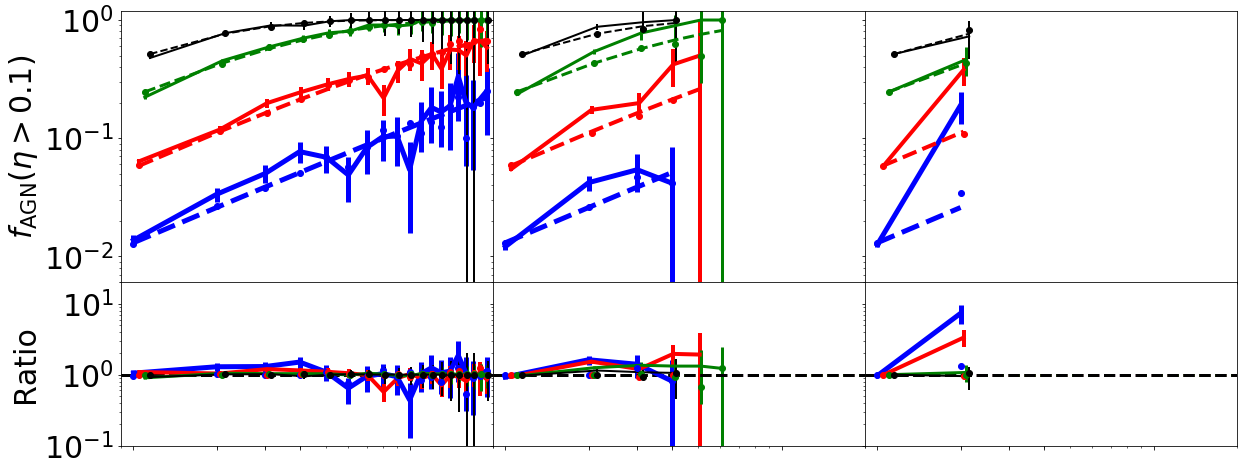}

\vspace{0.5 cm}

\includegraphics[width=13.5cm]{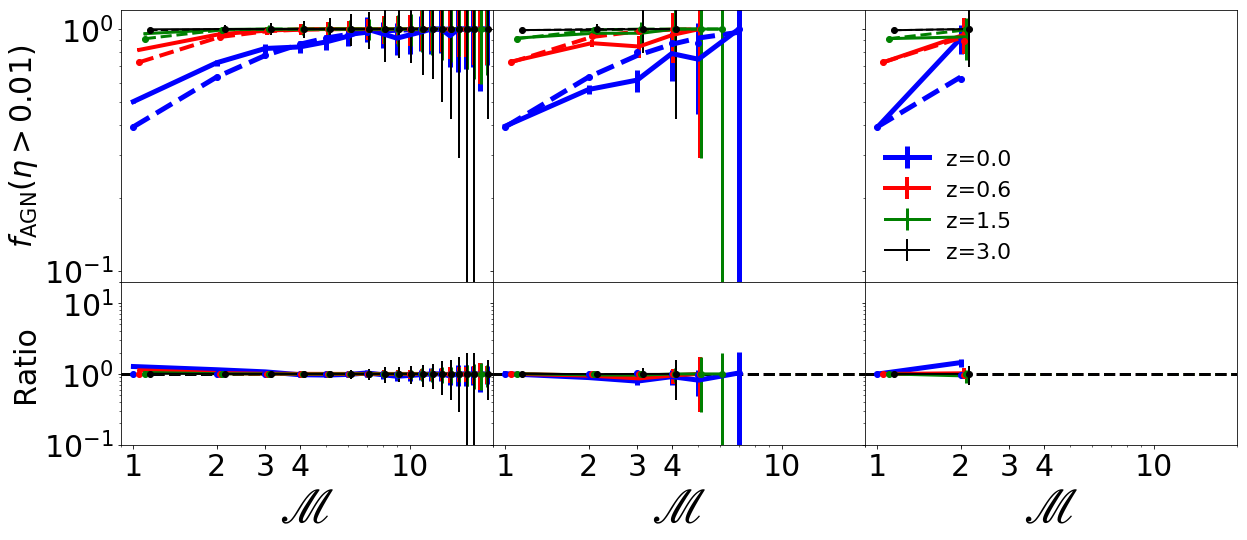}

\caption{\textbf{Upper/ larger panels:} The AGN fraction~($f_{\mathrm{AGN}}$) is defined as the fraction of BH systems that have at least 1 AGN. Each row assumes a different Eddington ratio threshold to define an AGN, as indicated in the y-axis labels. The solid lines are the AGN fractions for BH systems in IllustrisTNG, with error bars corresponding to Poisson errors in their number counts. The  filled circles show the median AGN fractions for 10 samples of randomly-selected BHs. 
Additionally, we also have dashed lines which also show predictions for the randomized systems, but computed analytically using Eq.~(5). As expected, the circles nearly overlap with the dashed lines. The left, middle, and right columns correspond to $d_{\mathrm{max}}=1$, 0.1, \& 0.01 $h^{-1}$ Mpc, respectively. \textbf{Lower/ smaller panels:} Ratios of AGN fractions with respect to analytical predictions for randomly-selected BH samples; 
i.e., the solid lines and filled circles are obtained by comparing 
the solid lines vs. dashed lines and circles vs. dashed lines, respectively, from the upper panels. At scales within $0.1$ and $0.01$ \Mpch, wherever adequate statistics are available, the highest Eddington ratio AGN ($\eta\geq0.1$) 
are more likely to be found in BH pairs, triples, and higher-order multiples than would be expected from random subsampling. The same is not true at scales of $\leq 1$ 
\Mpch, where the higher likelihood for multiple BH systems to contain at least one luminous AGN is mostly a result of simple combinatorics.  
For low Eddington ratio thresholds~($\eta>0.01$), AGN fractions are always high, such that little to no enhancement of AGN activity is detectable in spatially-associated BH systems.}

\label{AGN_fractions}
\end{figure*}

\subsection{AGN fractions of BH systems}
\label{AGN_fractions_of_black_hole_systems}
Figure \ref{AGN_fractions} shows the fraction~($f
_{\mathrm{AGN}}$) of BH systems~(as a function of multiplicity) that contain at least one AGN, where ``AGN" are defined via various Eddington ratio thresholds. We see that across the entire range of Eddington ratios and redshifts, systems with higher multiplicity 
have higher AGN fractions. 
However, this is to some extent a trivial consequence of the higher probability that a system containing many BHs will contain at least one AGN. In order to quantify the enhancement in AGN activity that can be attributed to environment, we
compare this result to AGN fractions~(filled circles in Figure \ref{AGN_fractions}) of the randomly-selected samples of BH systems~(see Section \ref{randomized_samples}). Additionally, we can also compute the AGN fractions of the randomized samples analytically (dashed lines in Figure \ref{AGN_fractions}). For a sample containing $N_{\mathrm{BH}}$ BHs and $N_{\mathrm{AGN}}$ AGN, 
the fraction~($f^{\mathrm{random}}_{\mathrm{AGN}}$) of randomly chosen sets of $\mathscr{M}$ BHs which have at least one AGN is given by 
\begin{equation}
    f^{\mathrm{random}}_{\mathrm{AGN}}=\frac{\binom{N_{\mathrm{BH}}}{\mathscr{M}}-\binom{N_{\mathrm{BH}}-N_{\mathrm{AGN}}}{\mathscr{M}}}{\binom{N_{\mathrm{BH}}}{\mathscr{M}}}.
    \label{eqn:AGN_fraction_randomized}
\end{equation} 
We see that the AGN fractions for the randomized samples computed using the two methods~(filled circles vs dashed lines) are consistent with each other, providing further validation for our use of randomized samples to identify true enhancements of AGN activity in spatially-associated BH systems. 

Also evident in Figure \ref{AGN_fractions} is that for BH systems on 1 \Mpch\ scales, there is no significant enhancement of AGN fractions compared to their corresponding randomized samples, even at the highest Eddington ratios ($\eta \geq 0.7$). 
Enhanced AGN fractions are seen for high Eddington ratio AGN in closer BH systems, however, suggesting a merger-triggered origin for these luminous AGN. The
following paragraphs describe more details about these enhancements at various Eddington ratio thresholds.   

Let us first focus on the AGN fractions of the most luminous AGN ($\eta \geq 0.7$). The topmost panels in Figure \ref{AGN_fractions} contain only data at $z\geq 1.5$ 
(at $z\leq0.6$, the very few $\eta \geq 0.7$ AGN that exist are insufficient to make statistically robust predictions). 
We see that at scales of 0.1 \Mpch,  
the AGN fractions of BH pairs and multiples~($\mathscr{M}>1$) are enhanced 
by up to factors of $\sim3-6$ compared to their corresponding randomized samples. At the smallest ($\leq 0.01$ \Mpch) scales, we see hints of a similar trend, but there are too few luminous AGN systems to draw definite conclusions. 
In contrast, 
BHs that are isolated ($\mathscr{M}=1$) at 
1 \Mpch\ scales 
are actually slightly \textit{less} likely to host luminous AGN than individual BHs sampled randomly from the overall population.  
These trends imply a strong association between luminous AGN triggering and BH multiplicity on $\leq 0.1$ \Mpch\ scales.

If the AGN Eddington ratio threshold is decreased to $\eta_{\rm min} = 0.1$, 
enhanced AGN fractions are seen for $d_{\rm max} \leq 0.1$ Mpch\ systems at all redshifts (excepting $z=3$, where most primary BHs are $\eta\geq0.1$ AGN even in the randomized samples). 
These AGN enhancements 
are smaller~(up to factors $\sim2$) than those for the most luminous AGN ($\eta>0.7$) at $\leq0.1$ \Mpch~  scales. On the smallest ($\leq0.01$ \Mpch) scales, however, the AGN fractions of BH pairs at low redshift 
are more strongly enhanced~(up to factors of $\sim 8$ at $z=0$) 
compared to their corresponding randomized pairs. 

If the AGN Eddington ratio threshold is further decreased to $\eta_{\rm min}=0.01$, no significantly enhanced AGN fractions 
are seen in any multiple BH systems. A notable excpetion is 
$\leq0.01$ \Mpch\ scale pairs at $z=0$, 
which are enhanced by up to factors of $\sim2$.


In a nutshell, the above trends indicate that AGN are more likely to be found in multiple BH systems at 1) high Eddington ratios 
and 2) smaller BH separations. Because multiple BH systems on $\leq 0.1$ \Mpch\ scales are likely to be hosted in ongoing galaxy interactions or mergers, this finding is in agreement with previous studies indicating enhanced AGN activity in interacting galaxies \citep{2011ApJ...737..101L,2011ApJ...743....2S,2011MNRAS.418.2043E,2013MNRAS.435.3627E,2014AJ....148..137L,2014MNRAS.441.1297S,2017MNRAS.464.3882W,2018PASJ...70S..37G,2019MNRAS.487.2491E}. Our findings are also 
in agreement with previous studies suggesting that luminous AGN are strongly clustered in rich environments at small scales~$\lesssim0.1$ \Mpch, in particular the ``one-halo" term of the AGN/ quasar clustering measurements~\citep{2012MNRAS.424.1363K,2017MNRAS.468...77E}.
Finally, the fact that no enhancements are seen at $1$ \Mpch\ scales is also consistent with previous clustering studies, which find no significant luminosity dependence on large scale clustering amplitude~\citep{2006MNRAS.373..457L,2018MNRAS.474.1773K,2019MNRAS.483.1452W,2020ApJ...891...41P}.   

\begin{figure*}
\centering
\includegraphics[width=14cm]{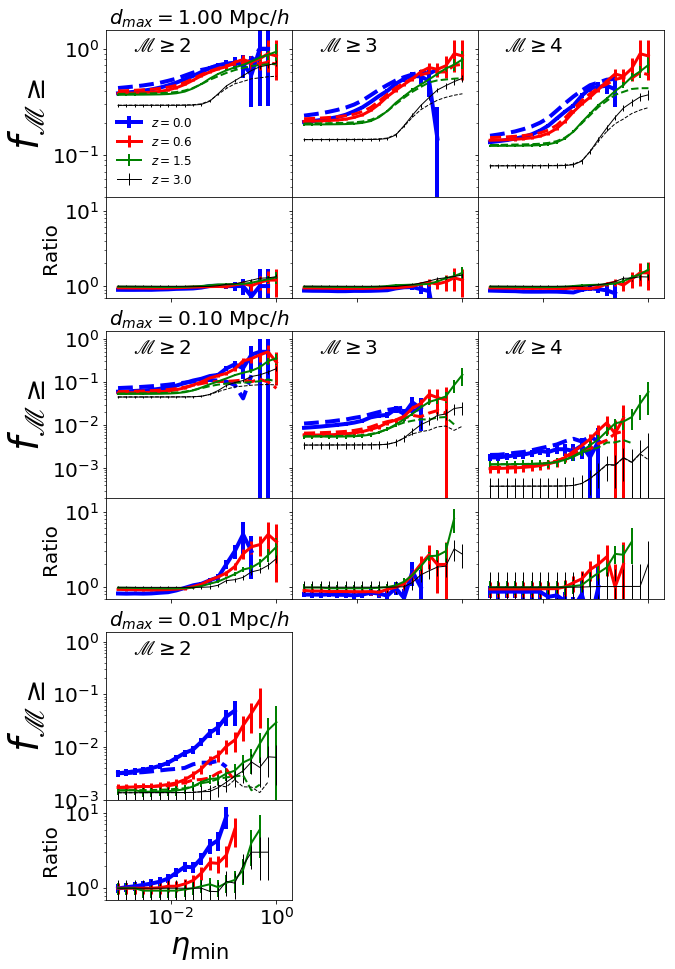}
\caption{\textbf{Upper/ larger panels}: $f_{\mathrm{\mathscr{M}(\geq)}}$ refers to the fraction of primary AGN that live in BH systems with a minimum multiplicity~$\mathscr{M}(\geq)$; this is plotted as a function of the threshold Eddington ratio $\eta_{\mathrm{min}}$. Solid lines correspond to the BH systems, and 
dashed lines correspond to the median values for the 10 samples of randomly-selected BHs. The top, middle, and bottom rows correspond to $d_{\mathrm{max}}=1$, 0.1, \& 0.01 \Mpch, respectively. The errorbars correspond to Poisson errors. Within each row, the \textbf{lower/ smaller panels} denote the ratio~(solid/ dashed lines) between the predictions for the true BH systems vs. that of the randomized systems. The left, middle and right panels correspond to systems with at least 2, 3 and 4 members, respectively. The different colors correspond to different redshifts. At higher Eddington ratios, the likelihood of AGN to belong to multiple BH systems within $0.1$ and $0.01$ \Mpch\ scales is significantly enhanced compared to that of the randomized systems; on the other hand, there is very little enhancement for BH systems within $1$ \Mpch\ scales.} 

\label{AGN_that_are_systems}
\end{figure*}

\begin{figure*}
\begin{tabular}{cc}
\includegraphics[width=8.1cm]{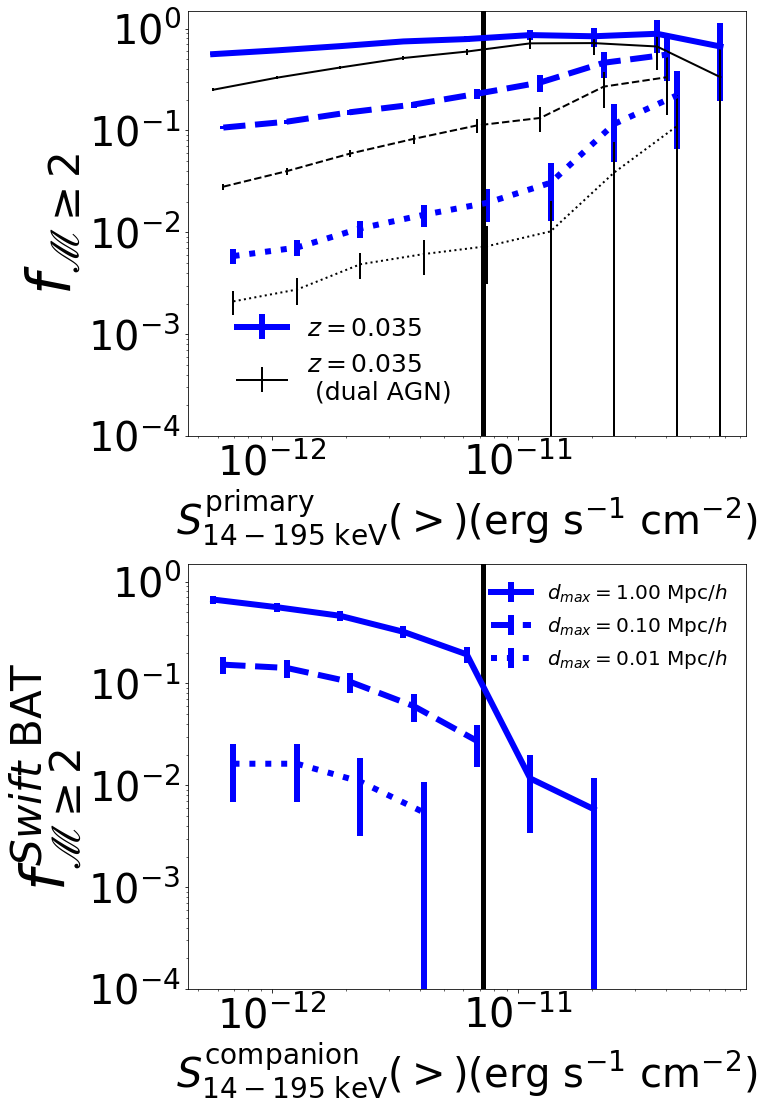} &

\includegraphics[width=8cm]{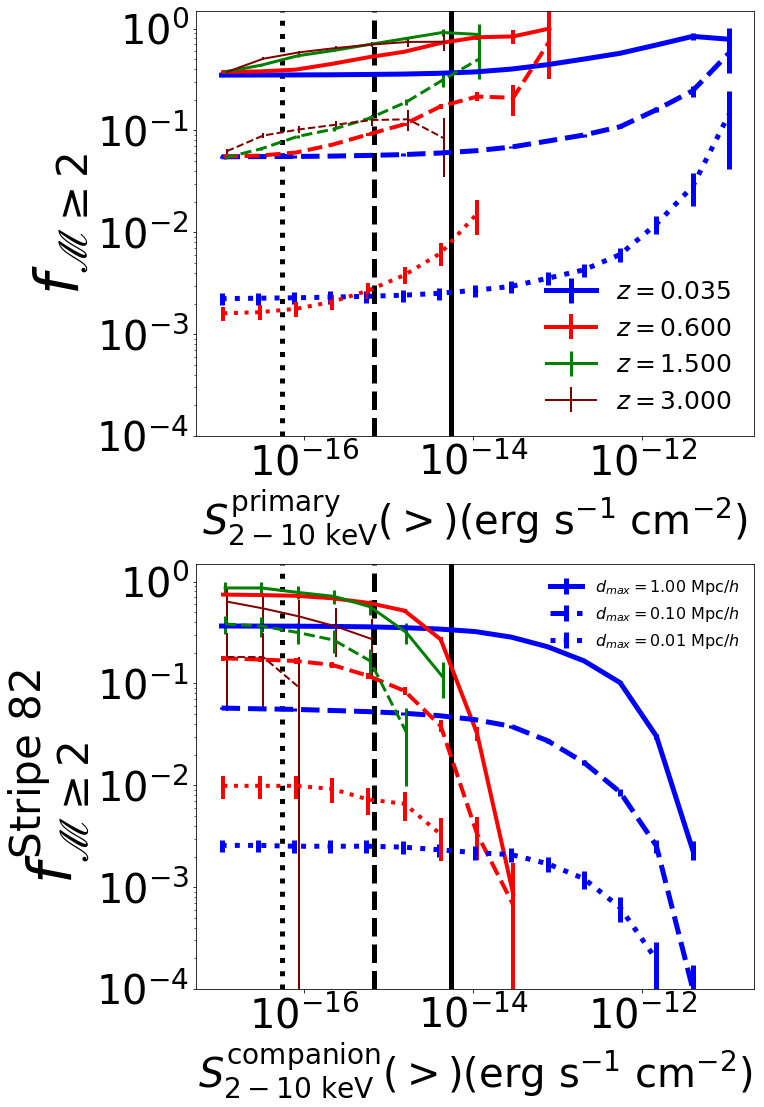} 

\end{tabular}
\caption{ \textbf{Upper panels:} The fraction $f(\mathscr{M}\geq2)$ of primary AGN~(defined here to be the most luminous member of the BH system) with 14-195 keV~(right panel) and 2-10 keV~(left panel) threshold flux, 
that live in pairs/ multiples. The different colors generally correspond to different redshifts~(see legend). In the left panel, the thin black lines correspond to systems where the companion BHs have $L_{2-10~\mathrm{keV}}>10^{42}~\mathrm{erg~s^{-1}}$ at $z=0.035$. Solid, dashed and dotted lines correspond to $d_{\mathrm{max}}=1$, 0.1, \& 0.01 $h$\inv\ Mpc, respectively. The assumed bolometric corrections for the X-ray bands have been adopted from \protect\cite{2010MNRAS.402.1081V}~(left panel) and \protect\cite{2012MNRAS.425..623L}~(right panel). The vertical line in the right panel marks the detection limit~($7.2\times10^{-12}~\mathrm{erg~s^{-1}~cm^{-2}}$) for the 105 month Swift-BAT survey~\protect\citep{2018ApJS..235....4O}. The vertical lines in the left panel mark detection limits of the following fields obtained from the \textit{Chandra} X-ray observatory: the dotted line~($5.5\times10^{-17}~\mathrm{erg~s^{-1}~cm^{-2}}$) corresponds to the \textit{Chandra} Deep Fields-North and South~(CDF-N and CDF-S); the dashed line~($6.7\times10^{-16}~\mathrm{erg~s^{-1}~cm^{-2}}$) corresponds to the Extended \textit{Chandra} Deep Field - South~(ECDF-S); and the solid line~($5.4\times10^{-15}~\mathrm{erg~s^{-1}~cm^{-2}}$) corresponds to the \textit{Chandra} Stripe 82 ACX survey~\protect\citep{2013MNRAS.432.1351L}. At the detection threshold of \textit{Swift}-BAT, the fraction of ultra-hard X-ray AGN associated with BH pairs and multiples within scales of $0.1$ \Mpch\ is $\sim10-20\%$. At the Stripe 82 ACX detection limit, the fraction of hard X-ray AGN in $0.1$\Mpch\ 
scale pairs and multiples is $\sim10-20\%$~($5\%$) at $z\sim0.6-1.5$~(at $z\sim0.035$).} \textbf{Lower panels:} $f^{\mathit{Swift}~\mathrm{BAT}}_{\mathrm{\mathscr{M}\geq2}}$~($f^{\mathrm{Stripe~ 82}}_{\mathrm{\mathscr{M}\geq2}}$) is defined as the fraction of \textit{Swift}-BAT~(Stripe 82) primary AGN that have at least one detectable companion above a given flux threshold ($x$-axes show the flux threshold of the companion AGN). $\sim3\%$ of \textit{Swift} BAT AGN at $z=0.035$ have companions within $0.1$ \Mpch~that are detectable at the 105-month survey limit.


\label{hard_xray}
\end{figure*}

\subsection{What fraction of observable AGN are members of BH systems?}
As mentioned earlier, some observational studies also look for `merger fractions' of AGN ---i.e., what fraction of AGN are hosted by merging/ interacting systems. In terms of our work, a robust proxy for these 
merger fractions is the fraction of AGN that are members of BH pairs and multiples. Figure \ref{AGN_that_are_systems}~(solid lines) shows the fraction~($f_{\mathscr{M}\geq}$) of primary AGN that are members of BH systems of a threshold multiplicity $\mathscr{M}\geq$, plotted as a function of threshold Eddington ratio $\eta_{\mathrm{min}}$. These are compared to corresponding predictions for AGN belonging to the randomized samples. We find that in the regime of Eddington ratio thresholds between $\sim0.01-1$, higher Eddington ratio AGN are more likely to have one or more companions compared to lower Eddington ratio AGN. However, this trend is also seen for the randomized samples which shows that this is~(in part) due to our choice of the most luminous member as the primary; this accompanies an inherent statistical bias in favor of more luminous AGN being more likely to be picked out from higher order BH systems. For $\eta<0.01$, $f(\mathscr{M}\geq)$ tends to gradually flatten for both randomized samples and true samples, but this is simply because as we continue to decrease the Eddington ratio threshold, we eventually cover the full AGN population. If we look at the redshift evolution at a fixed Eddington ratio threshold, we see that lower-redshift AGN have a higher probability of being a member of a BH multiple. This is primarily because at fixed multiplicity, BH systems tend to have decreasing Eddington ratios at lower redshifts due to a general  decrease in AGN luminosity with decreasing redshift~(as seen in Appendix \ref{appendix1}: Figure \ref{appendix1_fig}); as a natural corollary, at fixed Eddington ratio, BH systems have higher multiplicities at lower redshifts.
We now look at the difference in the $f(\mathscr{M}\geq)$ between the true samples of BH systems and the corresponding randomized samples in order to filter out the effects that are physical~(see ratio plots of Figure \ref{AGN_that_are_systems}). At scales of $1$ \Mpch, we see no significant difference between $f(\mathscr{M}\geq)$ predictions for the true samples and the randomized samples for the entire range of Eddington ratio thresholds; this is similar to our findings for the AGN fractions in the previous section. As we approach scales of $0.1$ \Mpch, we find that $f(\mathscr{M}\geq)$ is enhanced for the true samples as compared to the randomized samples at high enough Eddington ratios. These enhancements start to appear at $\eta \sim 0.01$ and increases up to factors of $\sim4$ for the most luminous AGN~($\eta\sim0.7-1$). At scales within $0.01$ \Mpch, we see the strongest enhancements; in particular, if we look at $z=0,0.6$ pairs where we have the best statistics, the enhancements are up to factors of $\sim7-9$ for the most luminous AGN~($\eta\sim0.1$). Furthermore, at $z=0$ the enhancements start appearing at Eddington ratios as low as $\eta\gtrsim0.001$. 

To summarize the above trends, we find that more luminous AGN have enhanced likelihood of having companion BHs within 
0.1 \Mpch; at the same time, there is no enhancement in the likelihood of AGN having companion BHs within $1$ \Mpch. This further corroborates the inferences drawn in the previous sections; i.e., no signatures of large scale AGN clustering are seen in 
our identified BH systems, but enhanced AGN activity is 
associated with rich environments small scales~($\leq0.1$ \Mpch), likely triggered by mergers and interactions between galaxies.  

\subsubsection{Companions of X-ray Selected AGN}

From an observational perspective, it is also instructive to 
estimate the fraction of AGN in BH pairs and multiples that would be detectable in a survey with a given flux limit. 
Therefore, in addition to analysing AGN samples characterized by Eddington ratios, we now repeat our analysis by characterizing AGN in multiple BH systems based on their estimated intrinsic fluxes in the 2-10 keV~(hard) and 14-195 keV~(\textit{Swift}/BAT ultra hard) X-ray bands (Figure \ref{hard_xray}, upper panels). For the 2-10 keV band, the bolometric corrections are adopted from \cite{2012MNRAS.425..623L}, where they assume best fit relations between the bolometric luminosities and 2-10 keV X-ray luminosities of the AGN samples from the \textit{XMM}-COSMOS~\citep{2009A&A...497..635C} survey. For the 14-195 keV X-ray band, we assume a constant bolometric correction of 15, as in previous 
analyses of \textit{Swift}/BAT AGN~\citep[e.g.,][]{2010MNRAS.402.1081V, 2012ApJ...746L..22K}. 
Figure \ref{hard_xray}~(upper panels) shows the fraction of AGN found in multiple BH systems ($f_{\mathscr{M}\geq2}$) as a function of the assumed hard or ultra-hard X-ray flux threshold. 
We see that in either case, brighter AGN are more likely to live in pairs and multiples, which is not surprising given the trends seen with Eddington ratios. 

We first focus on predictions for ultra-hard X-ray AGN (Figure \ref{hard_xray}, left panels). The 105-month all-sky \textit{Swift}/BAT survey has a flux limit of $7.2\times10^{-12}~\mathrm{erg~s^{-1}~cm^{-2}}$ in the 14-195 keV band \citep{2018ApJS..235....4O}. At these X-ray energies, even heavily obscured AGN experience little attenuation. Coupled with its sky coverage, this means that the \textit{Swift}/BAT survey yields a uniquely complete sample of low-redshift AGN ($z\lesssim0.05$). We present predictions for the simulation snapshot at $z=0.035$~(maroon lines in Figure \ref{hard_xray}: upper-left panels).  
Let us first look at $1$ \Mpch\ scales, where we previously found no enhanced AGN activity in BH pairs or multiples. We see that the majority~($\sim70-80\%$) of the detectable AGN live in BH pairs and multiples within scales of $1$ \Mpch. In line with our previous results, however, we conclude that this is primarily driven by the gravitational clustering of halos hosting BHs and has little to do with the AGN activity of the BHs. On scales $\leq 0.1$ \Mpch, where we did previously find enhanced AGN activity in multiple BH systems, we see that $\sim10-20\%$ of the detectable AGN population is associated with BH pairs and multiples. 

We can compare this population to the \textit{Swift}/BAT-selected AGN sample studied in \citet{2012ApJ...746L..22K}. Using optical imaging, they selected BAT AGN hosted in galaxies that have companions within 100 projected kpc. \citet{2012ApJ...746L..22K} then compared with 2-10 keV X-ray observations to identify those companions that also hosted AGN to determine the dual AGN frequency on these scales. They assumed a minimum AGN luminosity of  $L_{2-10~\mathrm{kev}} > 10^{42}$ erg~s\inv, to avoid confusion with X-ray emission from star-forming regions. We apply similar criteria to identify dual AGN in our data, selecting AGN that would be detectable in the 105-month BAT survey and that have companion AGN within 0.1 \Mpch\ with $L_{2-10~\mathrm{kev}} > 10^{42}$ erg s\inv. Using these criteria, we find that $\sim10\%$ of BAT-detected AGN in our sample are dual AGN on 0.1 \Mpch\ scales (thin lines in Figure \ref{hard_xray}, upper left panel). This is consistent with the results of \citet{2012ApJ...746L..22K}.


In the lower left panel of Figure \ref{hard_xray}, we examine the fraction of BAT AGN with at least one companion BH that would also be detected at the limit of the BAT survey. The fraction decreases with increasing flux, owing to the rarity of luminous AGN. We see that at $z\sim0.035$, $\sim3\%$ and $\sim10\%$ of BAT AGN have companions within $0.1$ and $1$ \Mpch, respectively, that are detectable at the 105-month survey limit. 

We similarly examine the companions of hard X-ray selected AGN, based on their inferred intrinsic 2-10 keV flux. We do not attempt to model the amount of AGN obscuration, although we note that many AGN have significant attenuation in the 2-10 keV band, particularly in late stage mergers~\citep[e.g.,][]{2015ApJ...814..104K,2017MNRAS.468.1273R,2020arXiv200601850S}. These results will therefore be most useful for comparison with X-ray AGN for which intrinsic luminosities can be estimated. We focus on the \textit{Chandra} Stripe 82 ACX survey~(solid vertical line in Figure \ref{hard_xray}), owing to its large area ($\sim17~\mathrm{deg^2}$) that yields statistically large samples of X-ray bright AGN. At $1$ \Mpch \ scales (due to gravitational clustering), $\sim70-80\%$ of the detectable AGN live in BH pairs and multiples at $z\sim0.6-3$; this decreases to $\sim40\%$ at $z\sim0.035$. At $\leq 0.1$ \Mpch\ scales, $\sim10-20\%$ of the detectable AGN population is associated with BH pairs and multiples at $z\sim0.6-1.5$; this decreases to $\sim5\%$ at $z\sim0.035$. Within $0.01$ \Mpch \ scales, $\lesssim1\%$ of the detectable AGN are associated with BH pairs. 

These findings are consistent with our previous results; for the Eddington ratio selected AGN in Figure \ref{AGN_that_are_systems}, we see that at the highest Eddington ratios, only up to $\sim40\%$ of AGN have companions within $0.1$ \Mpch. Overall, this suggests that the majority of AGN activity is actually \textit{not} associated with mergers/ interactions, but is instead driven by secular processes. This agrees with other recent observational~\citep{2014MNRAS.439.3342V,2017MNRAS.466..812V,2019ApJ...882..141M,2019ApJ...877...52Z} and theoretical studies~\citep{2020MNRAS.494.5713M}.

Finally, we examine the fraction of Stripe 82 AGN that would have companions detectable at various flux limits, giving rise to dual or multiple AGN~(lower right panel of Figure \ref{hard_xray}). Here we primarily focus on summarizing the results for companions within $\leq0.1$ \Mpch~(where we report statistically robust evidence of AGN enhancements), but results at $\leq0.01,1$ \Mpch ~are also presented in  Figure \ref{hard_xray} for completeness. At $z\sim0.035$, where the Stripe 82 flux limit corresponds to an AGN luminosity of $\sim 3 \times 10^{40}$ erg s\inv, almost all the available companions are detectable, implying that $\sim5\%$ of Stripe 82 AGN have companions already detectable without deeper observations. However, such low luminosities are quite difficult to distinguish from X-ray emission from star-forming regions. At $z\sim0.6$, where even intensely star forming regions are unlikely to mimic detectable AGN at the Stripe 82 limit, $\sim2\%$ of Stripe 82 AGN have companions detectable without deeper observations. At higher redshifts~($z\sim1.5,3$), there are no companions that are detectable within Stripe 82. However, the prospect of detecting companions is better for deeper observations such as the \textit{Chandra} deep field~(CDF) and extended \textit{Chandra} deep field~(ECDF) surveys. In particular, at the flux limit of the ECDF, almost all the available companions are detectable, implying that $\sim20\%$ \& $\sim30\%$ of Stripe 82 AGNs at $z=1.5$ \& $z=3$, respectively, have companions detectable at the ECDF limit.

\subsection{Disentangling AGN enhancements in multiples from trends with host mass}
\label{decoupling_degeneracy}
\begin{figure*}
\centering
\includegraphics[width=14.7cm]{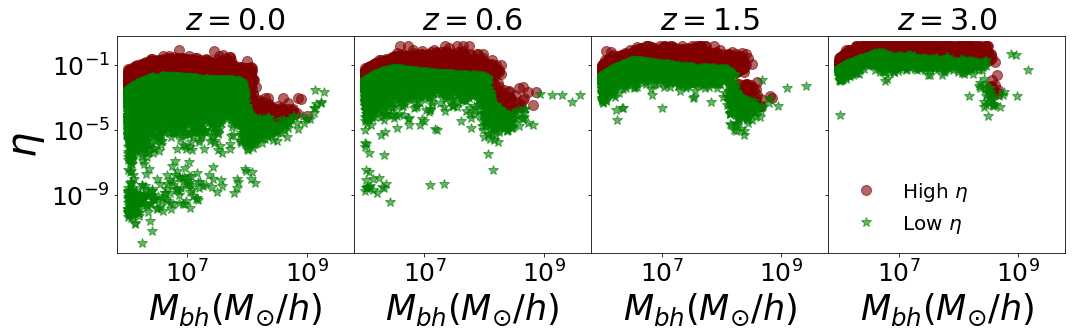}
\caption{AGN Eddington ratio~($\eta$) vs. BH mass~($M_{\rm bh}$). The overall sample is split into objects with Eddington ratios higher~(maroon color) and lower~(green color) than the median Eddington ratio at fixed BH mass. 
From left to right, the panels show snapshots at $z=$ 0, 0.6, 1.5 \& 3.  
This demonstrates how we split the BHs 
into high- and low-Eddington ratio populations 
to investigate the relative likelihood of these populations
to live in BH pairs and multiples.}
\label{AGN_selection}

\includegraphics[width=14.7cm]{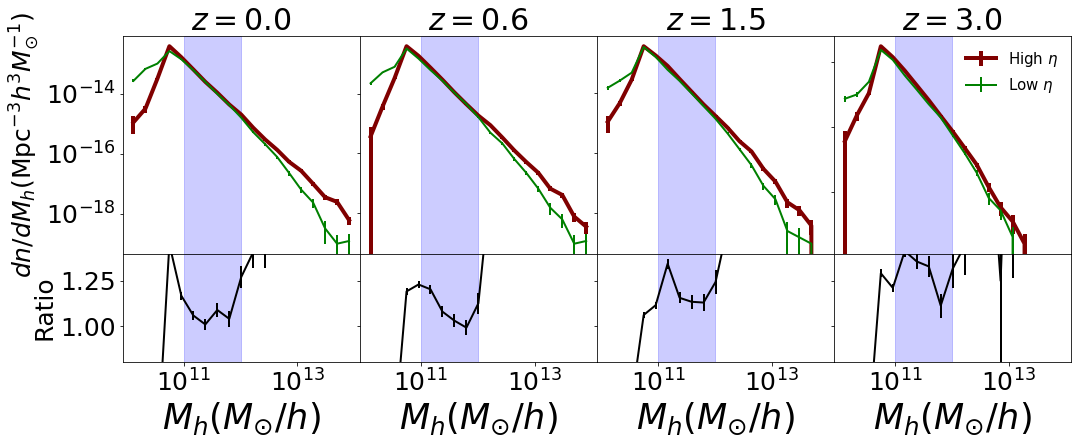}
\caption{\textbf{Top panels}: Maroon and green solid lines are the host halo mass~($M_{h}$) functions of BHs with Eddington ratios higher and lower than the median value at fixed BH mass. \textbf{Bottom panels}: Ratio between the blue vs. red lines presented on the top panels. The green region ($10^{11}<M_{h}<10^{12}~h^{-1}~M_{\odot}$) represents the BHs that 
were selected for the computation of $f(\mathscr{M}\geq)$ in Figure \ref{quartiles}. 
This region is chosen to ensure that the halo mass functions for the high- and low-Eddington ratio 
populations of Figure \ref{AGN_selection} match to within $\sim30\%$. From left to right, the panels show 
snapshots at $z=$ 0, 0.6, 1.5 \& 3.}
\label{host_halo_mass_functions}

\includegraphics[width=14.7cm]{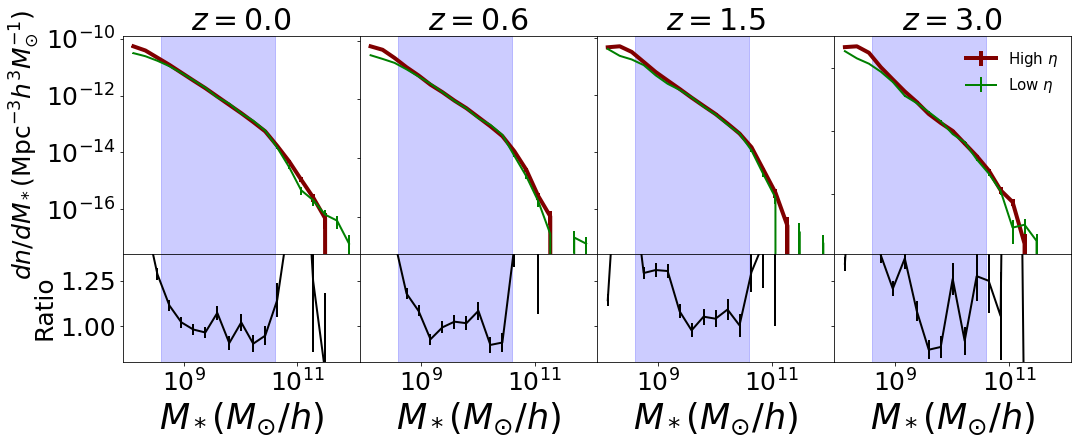}
\caption{\textbf{Top panels}: Blue and red solid lines are the host galaxy~(subhalo) stellar mass~($M_*$) functions of BHs with Eddington ratios higher and lower than the median value at fixed BH mass. \textbf{Bottom panels}: Ratio between the blue vs. red lines presented on the top panels. The green region represents the BHs that 
were selected for the computation of $f(\mathscr{M}\geq)$ in Figure \ref{quartiles}. This region of $10^{8.6}<M_{h}<10^{10.6}~h^{-1}~M_{\odot}$ is chosen 
to ensure that the stellar mass functions for the high- and low-Eddington ratio 
populations of Figure \ref{AGN_selection} match to within $\sim30\%$. From left to right, the panels show snapshots at $z=$ 0, 0.6, 1.5 \& 3.} 
\label{host_stellar_mass_functions}

\end{figure*}

\begin{figure*}
\centering
\includegraphics[width=\textwidth]{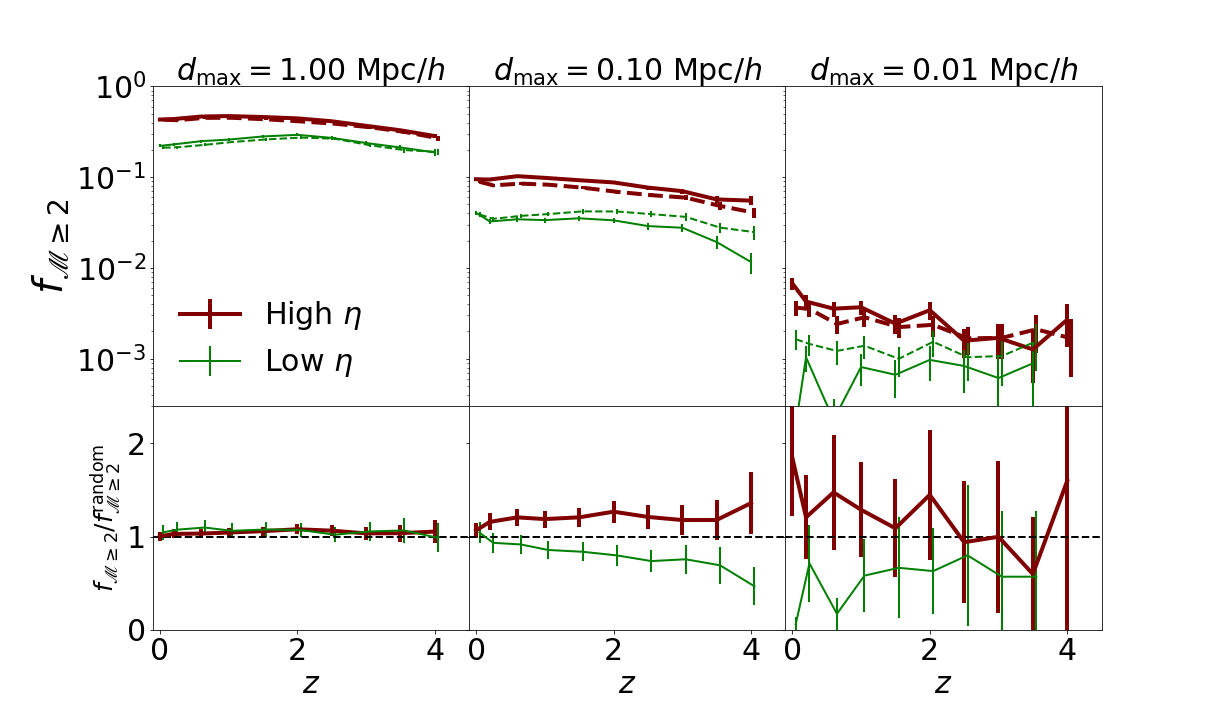}
\caption{\textbf{Upper/ larger panels:} $f(\mathscr{M}\geq2)$ as a function of redshift is the fraction of primary AGN that live in pairs. The blue and red lines represent primary AGN with Eddington ratios higher and lower, respectively, than the median value at fixed BH mass. The solid lines correspond to true BH systems, while the 
dashed lines correspond to the median values of 10 randomized systems. 
We further subsample the populations such that host halo and stellar masses 
are confined to the green highlighted regions in Figures \ref{host_halo_mass_functions} and \ref{host_stellar_mass_functions}, respectively. \textbf{Lower/ smaller panels:} The ratio 
between the predictions of $f(\mathscr{M}\geq)$ for the true samples of BH systems vs. that of the randomized samples. We find that the high Eddington ratio primary AGN~(blue color) have \textit{slightly higher likelihood} of belonging to multiple BH systems within scales of $0.1$ and $0.01$ \Mpch, compared to 
random subsets of BHs. At the same time, low Eddington ratio primary AGN~(red color) have \textit{slightly lower likelihood} of belonging to a multiple BH systems at scales within $0.1$ and $0.01$ \Mpch, compared to randomly chosen subsets of BHs. These effects are not seen at scales of $1$ \Mpch.}
\label{quartiles}
\end{figure*}

We have so far established that there is enhanced AGN activity in close systems of multiple BHs ($\leq 0.1$ \Mpch). The AGN enhancement at these scales may partly be attributed to AGN triggering by galaxy mergers and interactions. At the same time, there may also be a contribution from: 1) a possible correlation between AGN Eddington ratio and the mass of its host halo or galaxy, and 2) the fact that more massive haloes host a higher number of 
galaxies containing BHs, and therefore are richer in both active and inactive BHs. 
In this section, we shall statistically control for the host halo and galaxy mass and further look for possible enhancements in the AGN activity within BH systems that can be solely attributed to small scale galactic dynamics. 



Figure \ref{AGN_selection} shows the Eddington ratio vs. the BH mass 
of the overall BH populations within \texttt{TNG100}. We first divide the population based on whether the luminosities are higher~(blue circles) or lower~(red circles) than the median Eddington ratio at fixed BH mass. We then ensure that for each of the two populations, we select subsamples with similar host halo masses and host galaxy stellar masses, using the following procedure. In Figure \ref{host_halo_mass_functions}, blue and red solid lines show the resulting host halo mass functions of the BH samples represented by circles of the corresponding color in Figure \ref{AGN_selection}. The sharp drop in the halo mass function at $M_h\sim5\times10^{10}~h^{-1}~M_{\odot}$ corresponds to the threshold halo mass for inserting BH seeds. (The small tail of $M_h<5\times10^{10}~h^{-1}~M_{\odot}$ halos correspond to those that have seeded BHs at an earlier time, but have lost some mass due to tidal stripping.) We find that for halos with $M_h\gtrsim10^{12}~h^{-1}~M_{\odot}$ and $M_h\lesssim10^{11}~h^{-1}~M_{\odot}$, the halo mass functions for hosts of high- and low-Eddington ratio BHs differ slightly in their normalization. Thus, for this analysis we focus on BHs hosted in $10^{11}< M_h < 10^{12}~h^{-1}~M_{\odot}$ halos at all snapshots between $0<z<4$, where the difference in the halo mass function is small~($\lesssim30\%$). 

We follow the same procedure for host galaxy stellar masses~($M_*$), wherein we select galaxies with $10^{8.6}<M_*<10^{10.6}~h^{-1}~M_{\odot}$ such that 
the stellar mass functions differ by $\lesssim30\%$ between high- and low-Eddington ratio BH hosts (Figure \ref{host_stellar_mass_functions}: green regions). 
Overall, we have 1) divided the BH population into those with Eddington ratios higher and lower than the median value at fixed BH mass, and 2) further selected subsamples of both the populations with minimal differences in their host halo masses and host galaxy stellar masses. In the process, we have constructed two BH subsamples with similar host halo properties that differ solely in their Eddington ratios. We can now quantify the fraction of AGN in each of these subsamples that live in BH pairs and multiples, relative to the corresponding randomized BH samples.

Figure \ref{quartiles}~(upper panels) shows the fraction $f(\mathscr{M}\geq2)$ of AGN that are primary members of BH pairs and multiples, 
plotted as a function of redshift for the high- and low-Eddington ratio 
populations of primary BHs described above (and in Figures \ref{AGN_selection}-\ref{host_stellar_mass_functions}). 
The solid and dashed lines correspond to the predictions for the true BH systems and the randomized samples, respectively. For both the randomized samples as well as true samples, more luminous AGN 
have a greater likelihood of being members of BH systems, compared to those that are less luminous. 
This, again, is in part due to the statistical bias arising due to our choice of the most luminous AGN as the primary. In order to isolate 
the physical effects, we look at the ratio of $f(\mathscr{M}\geq2)$ between the true samples and the randomized samples, which are shown in Figure \ref{quartiles}~(lower panels). At scales within $0.1$ \Mpch, we find that~(with the exception of $z\sim0$) the likelihood for more luminous AGN 
to live in BH pairs and multiples is enhanced for the true samples compared to that of randomized samples. Likewise, the likelihood of less luminous AGN 
to live in BH pairs and multiples is suppressed for the true samples compared to that of randomized samples. Therefore, at $z\gtrsim0.6$, we see clear 
evidence that at scales of $\leq0.1$ \Mpch, BH pairs and multiples 
are indeed associated with more enhanced AGN activity, independent of the overall masses of the host halos and galaxies. 

At $z\sim0$, 
we do not see any enhancement at $\leq0.1$ \Mpch\ scales. This is simply because 
the typical Eddington ratios at $z=0$ are lower overall~(median Eddington ratio $\sim0.01$, see Figure \ref{AGN_selection}). As we saw in Figure \ref{AGN_that_are_systems}, no significant enhancements are seen in low-luminosity ($\eta\lesssim0.01$) AGN for $d_{\rm max}\geq0.1$ \Mpch\ BH systems. 
However, if we look at $\leq0.01$ \Mpch\ scales, we do see evidence of enhanced AGN activity at $z\lesssim2$, including at $z=0$. 

Additionally, note that the enhancements in $f(\mathscr{M}\geq2)$ 
seen in Figure \ref{quartiles} are significantly smaller than the strongest enhancements reported for the most luminous AGN~($\eta\gtrsim0.7$) shown in Figure \ref{AGN_that_are_systems}. But this is simply because selecting the high Eddington ratio samples in Figure \ref{quartiles} is broadly equivalent to samples with ``effective Eddington ratio thresholds" ranging between $\eta\sim0.01$ at $z=0$ to $\eta\sim0.1$ at $z=3$~(see Figure \ref{AGN_selection}); these values are significantly smaller than $\eta\gtrsim0.7$ and therefore correspond to weaker enhancements. 
Lastly, we also show 
the results for the $\leq1$ \Mpch\ scales (leftmost panels in Figure \ref{quartiles}), wherein 
we find no difference in $f(\mathscr{M}\geq2)$ between true and randomized BH systems; 
this is expected, given the results in Figure \ref{AGN_that_are_systems}.       

To summarize, the results in this section further solidify 
the association of enhanced AGN activity with BH pairs and multiples within 
scales of 
0.1 \Mpch. When controlled for host halo mass, the fraction of high-Eddington-ratio AGN that are in multiple BH systems is only modestly enhanced over random associations. Nonetheless, our results demonstrate that this trend 
does exist independent of the fact that 
massive halos and galaxies tend to host more luminous and more numerous BHs. This enhancement in AGN activity is likely driven by mergers and interactions between galaxies.   



\subsection{Impact of small scale environment on the Eddington ratios of BH systems}
\label{impact_on_edd}
\begin{figure*}
\includegraphics[width=17cm]{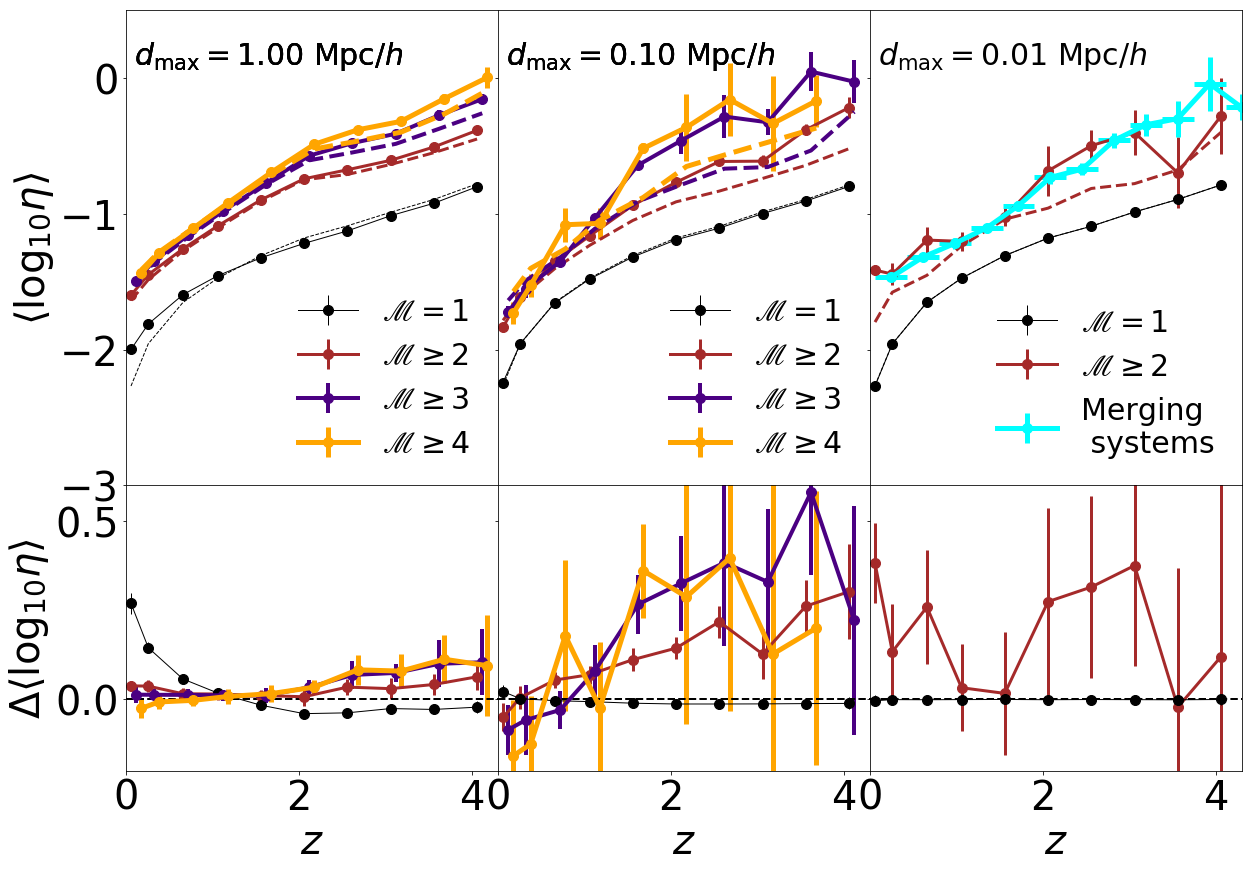}
\caption{\textbf{Upper / larger panels}: Median values of the AGN eddington ratio $\left<\log_{10}\eta\right>$ of the primary BHs of pairs and multiples as a function of redshift. The blue, red, and green circles correspond to all multiple BH systems with $\mathscr{M}\geq2,3,$ \& 4, respectively. The black circles correspond to isolated BHs~($\mathscr{M}=1$). The dashed lines correspond to the median values for 10 samples of randomly-selected systems. The cyan lines correspond to pairs of merging BHs recorded at high time resolution during the simulation run; the sample of merging BHs has 
been modulated to have similar distributions of masses and mass ratios as the $0.01$ \Mpch\ pairs. \textbf{Lower / smaller panels}: The difference~(solid$-$dashed lines) between 
$\left<\log_{10}\eta\right>$ for the true BH systems vs. 
the randomized systems. The different rows correspond to various values of $d_{\mathrm{max}}$. The error-bars on the y axis are obtained using bootstrap resampling. We see that Eddington ratios associated with primary BHs are enhanced when they belong to BH systems on 0.1 or 0.01 \Mpch\ scales, relative to 
randomly chosen subsets of BHs. On $0.01$ \Mpch\ scales, the AGN enhancements 
are most significant at $z\sim0$. In contrast, little enhancement in AGN activity is seen in 
BH systems at scales of $1$ \Mpch.} 
\label{average_eddington_ratio}
\end{figure*}

Having established the influence of small scale~($\leq 0.1$ \Mpch) environment on AGN activity, we now quantify in greater detail the magnitude of Eddington ratio enhancements in 
BH pairs and multiples compared to isolated BH.  In particular, we look at the Eddington ratios associated with the primary members of BH multiples~($\mathscr{M}\geq2,3,4$) as well as those 
of isolated BHs~($\mathscr{M}=1$), 
as a function of redshift. Figure \ref{average_eddington_ratio} shows that the Eddington ratios for the true sample of BH multiples 
increase with multiplicity at fixed redshift, but so do the Eddington ratios of the random samples. This again owes to the fact that 
the primary AGN form a biased sample compared to the full population. 

We are most interested in the comparison of the median Eddington ratios between the true samples of BH multiples and 
the randomized samples~(solid vs dashed lines); these are shown in the lower panels of Figure \ref{average_eddington_ratio}. At scales of $1$ \Mpch, the median Eddington ratios for the true samples of BH multiples have minimal differences~($\lesssim0.1$ dex) with respect to the randomized samples, as expected from our analysis so far. At scales of $d_{\mathrm{max}}=0.1$ \Mpch~(middle panel), we find that the Eddington ratios for the true samples of BH multiples tend to be increasingly 
enhanced 
at higher redshifts. At the highest redshifts of $z\sim3-4$, the enhancements are up to $\sim0.3-0.5~\mathrm{dex}$. This agrees with our results in Figure \ref{quartiles} and likely reflects 
stronger enhancements in merger-driven AGN activity due to greater availability of cold gas at higher redshifts. At the lowest redshifts of $z\lesssim0.6$, there is no significant enhancement of the Eddington ratios at $\leq0.1$ \Mpch\ scales. 

At 
separation scales of $d_{\mathrm{max}}=0.01$ \Mpch, however, 
where we only have BH pairs, 
we find that at $z\lesssim0.6$, the median Eddington ratios for the true samples are enhanced by $\sim0.3-0.4~\mathrm{dex}$. (At higher redshifts, some evidence of enhanced Eddington ratios is also seen, but the results are at best only marginally significant.) 
As we discussed earlier, our sample of BH pairs at these scales is incomplete, because a significant fraction of them merge prematurely due to the BH repositioning scheme implemented by the simulation. Therefore, we simultaneously look at the complete sample of BH mergers, which are recorded at much higher time resolution than the snapshot data. 
There are $15953$ merger events recorded during the simulation run. We modulate  this sample of merging BHs such that it has 
similar distributions of masses and mass ratios as 
the sample of $0.01$ \Mpch\ BH pairs. We perform the modulation based on randomly selecting subsamples of BHs at various bins of mass ratios and masses, where the relative fraction of objects in each bin is tuned to represent the mass ratio and mass distributions of $0.01$ \Mpch\ BH pairs~(we do this to make the samples more comparable, since the Eddington ratios have been found to depend on the mass ratios as well as masses of the merging BHs). After the modulation we end up with $1970$ merging BH systems 
(as compared to the much smaller number of $0.01$ \Mpch\ scale BH pairs, which is only 285). We find that the median Eddington ratios of these merging BH pairs~(cyan lines in the rightmost panel) are broadly consistent with that of the $0.01$ \Mpch\ scale pairs. Thus, we can conclude that the incompleteness of small-separation pairs does not introduce systematic bias into our results, and that the higher Eddington ratios seen in these close pairs reflect a genuine enhancement in AGN activity. 


To summarize, we find 
a measurable impact of the small scale~($\leq0.1$ \Mpch) environment on 
AGN Eddington ratios, 
which generally tends to increase at higher redshift. The median Eddington ratios are, at best, enhanced up to factors of $\sim2-3$~($0.3-0.5$ dex). This supports the existence of a merger-AGN connection. However, because the enhancements in AGN activity for BH pairs and multiples are relatively modest, our results do not suggest that merger-driven AGN fueling is a dominant channel of BH growth overall \citep[see also][and Thomas et.~al., in prep.]{2020MNRAS.494.5713M}

\section{Conclusions}
\label{conclusions}

In this work, we investigate the role of environment 
on 
AGN activity within the \texttt{TNG100} realization of the \texttt{Illustris-TNG} simulation suite. In particular, we investigate whether BH pairs and multiples~(within separations of $0.01-1$ \Mpch) 
have enhanced AGN activity compared to samples of randomly assigned pairs and multiples. 

The number density of BHs in TNG100 is 
$n \sim 0.06\,h^3$ Mpc$^{-3}$ at $z\lesssim1.5$ ($n\sim 0.02\, h^3$ Mpc$^{-3}$ at $z=3$). About $10\%$ of these BHs live in pairs on scales of 0.1 \Mpch, and $\sim10\%$ of these pairs (i.e., $\sim 1\%$ of all BHs) have additional companions, forming triples or higher-order multiples. A similar fraction ($\sim12\%$) of BHs are in pairs on 1 \Mpch\ scales, but $\sim30\%$ of these ($\sim3.6\%$ of all BHs) have additional companions on these scales. On the smallest scales ($d_{\rm max}=0.01$ \Mpch), in contrast, only $\sim 0.2\%$ of BHs are found in pairs (though as discussed above, this sample of pairs is incomplete). Overall,
pairs and triples live in haloes with a range of masses, but the median host halo mass ($\lesssim10^{12}~h^{-1}~M_{\odot}$) varies little with redshift. 


We find that the AGN activity associated with these BH systems is enhanced at scales within $0.01$ \Mpch\ and $0.1$ \Mpch\ across the entire redshift regime~($z\sim0-4$) we covered in this study. However, no such enhancements are found for BH systems within $1~\mathrm{Mpc}$ scales. The lack of enhancements in AGN activity at $\sim1~\mathrm{Mpc}$ scales, is consistent with recent observational constraints on large scale clustering, which were found to exhibit no significant dependence on AGN luminosity. On the other hand, the enhancements at smaller scales~$\sim0.01~\& 0.1$ \Mpch\ can be attributed to AGN activity triggered by merging and interacting galaxies. 


The influence of the small scale~($\leq0.1$ \Mpch) environment on the AGN activity is strongest at high Eddington ratios. 
In particular, for the highest Eddington ratio~($\gtrsim0.7$) AGN, the AGN fractions are significantly enhanced~(up to factors of $\sim3-6$) for pairs, triples and quadruples at scales within $\leq0.1 $ \Mpch\ compared to random BH samples. 
As we decrease the Eddington ratio thresholds, these environmental enhancements gradually become smaller and eventually disappear around Eddington ratios of $\sim 0.01$. 
Additionally, the enhancements~(at fixed Eddington ratio) also tend to be highest at the smallest ($\leq0.01$ \Mpch) scales. For example, at Eddington ratios greater than $0.1$, the AGN fractions of $\leq0.01$ \Mpch\ pairs at $z=0$ are enhanced up to factors of $\sim8$. Similarly, we also find that more luminous AGN have an enhanced likelihood~(up to factors of $\sim4$ and $\sim9$ within 0.1 and $0.01$ \Mpch\ scales, respectively) of living in BH pairs and multiples, compared to random subsamples of BHs.

In order to control for possible systematic biases, we investigate whether
our results are influenced by the possibility that more luminous AGN tend to 
live in more massive 
galaxies and halos, which incidentally tend to also host a higher number of BHs. We found that even after statistically controlling for the host halo mass and host galaxy stellar mass, more luminous AGN continue to have enhanced likelihood of living in BH pairs and multiples within 
0.1 \Mpch, compared to random subsamples of BHs. This further solidifies the 
correlation between AGN activity and the richness of the small scale~($\leq0.1$ \Mpch) environment over the entire redshift range between 0 to 4. Additionally, we find that the enhancement in accretion rates within BH systems is stronger at higher redshift, which presumably reflects the 
higher cold gas fractions at higher redshifts. 
 
Because the Eddington ratio of AGN is not a directly observable quantity (and BH mass measurements must often rely on indirect methods), we also estimate the X-ray luminosities of the AGN in our sample and determine the likelihood for X-ray selected AGN to live in BH pairs and multiples, as a function of the X-ray flux limits relevant to current surveys. At the limit of the 105 month \textit{Swift}-BAT survey, about $\sim10-20\%$ of detectable AGN at $z=0.035$ have at least one secondary companion within $0.1$ \Mpch \
scales. $\sim3\%$ of these BAT AGN have companions that are also detectable at the \textit{Swift}-BAT survey flux limit. Additionally, when we define dual AGN as in~\cite{2012ApJ...746L..22K}~(i.e., when AGN companions are selected based on a minimum 2-10 keV luminosity of $10^{42}$ erg s\inv), we report a dual AGN frequency of $\sim10\%$, consistent with their measurements. 

If instead we consider the companions of AGN selected in the 2-10 keV band at the limit of the \textit{Chandra} Stripe 82 survey (with no constraints on the ultra-hard X-ray band), we find that $\sim 5$\% of AGN live in pairs and multiples within 0.1 \Mpch \ scales at $z=0.035$. At higher redshifts~($z\sim0.6-1.5$), up to  $\sim30\%$ of such AGN have companions within 0.1 \Mpch\ scales. However, for only $\lesssim2\%$ of these $z\gtrsim0.6$ AGN, the companions are detectable without observations deeper than Stripe 82. At the flux limits of ECDF, most of the companions~(up to $z\sim3$) are available for detection, but those with low X-ray luminosities will likely be indistinguishable from star formation, and many will also have significant dust attenuation. 

With its wide-field imaging capabilities, the upcoming Advanced Telescope for High Energy Astrophysics (Athena) mission~\citep{2013sf2a.conf..447B} will enable new surveys that are expected to detect hundreds of AGN at $z>6$~\citep{2013arXiv1306.2307N}. The proposed Advanced X-ray Imaging Satellite (AXIS)~\citep{2018SPIE10699E..29M} and Lynx X-ray Observatory (Lynx) missions~\citep{2018arXiv180909642T} would enable detection of large new populations of AGN including high redshift AGN and close ($\lesssim 0.01$ \Mpch) dual AGN, owing to their sub-arcsecond imaging requirements and their factors of 10 and 100, respectively, better sensitivity than \textit{Chandra}. Our finding that merger-driven AGN activity is a significant but subdominant channel for BH fueling in TNG100 provides additional motivation for pursuing these key science goals with Athena.


While the enhanced AGN activity in rich, small-scale environments is 
consistent with the presence of the merger-AGN connection, we find 
that only a subdominant~(at best $\sim40\%$ for the highest Eddington ratio AGN) fraction of AGN actually live in BH pairs and multiples. Furthermore, enhancements in the Eddington ratios in BH pairs and multiples are, at best, only up to factors $\sim2-3$. Therefore, most AGN fueling as well as BH growth in \texttt{TNG100} may still be primarily triggered by secular processes, with a significant but minor role played by galaxy mergers/ interactions. We plan to explore this question in more detail in future work, including our companion paper, Thomas et.~al.~(in prep).

\appendix
\section{Scaling relation between Eddington ratio vs multiplicity of BH systems}
\label{appendix1}
\begin{figure*}
\includegraphics[width=18cm]{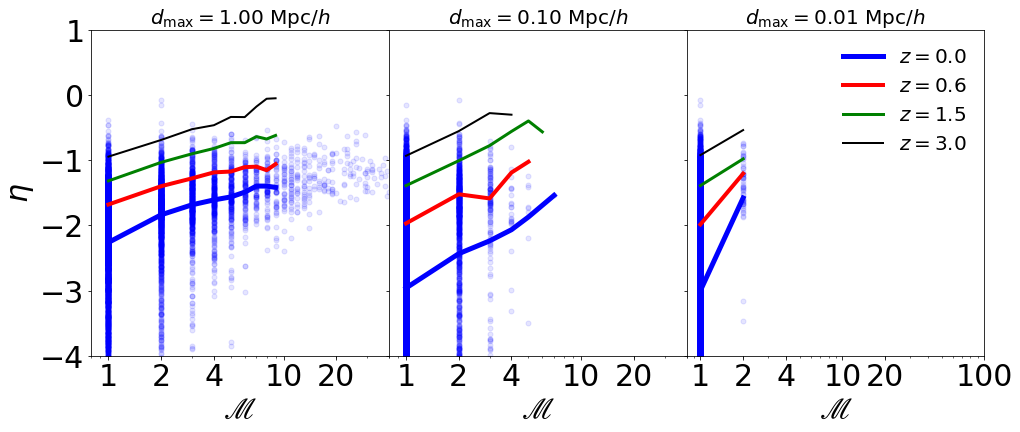}
\caption{Eddington ratio~($\eta$) as a function of the Multiplicity~($\mathscr{M}$) of the primary BH~(member with the highest Eddington ratio), for BH systems as predicted by the TNG100 simulation. The blue circles correspond to the scatter at $z=0$. The solid lines are the median relations at various redshifts. We see a positive correlation between the BH multiplicity and the Eddington ratio of the primary member, which tends to be stronger at smaller scales~(particularly at $z=0,0.6$). The redshift evolution shows that at fixed multiplicity, Eddington ratios decrease with decreasing redshifts; conversely, at fixed Eddington ratio, BH systems have higher multiplicities at lower redshifts.}
\label{appendix1_fig}
\end{figure*}

Here, we briefly discuss the overall scaling relation between Eddington ratio vs. BH multiplicity, which is shown in Figure \ref{appendix1_fig}. We see that BH systems having higher Eddington ratios~(for the primary member) have higher multiplicities at all scales between $0.01-1$ \Mpch. The correlation tends to be somewhat stronger at smaller separation scales, particularly at $z\sim0,0.6$. The redshift evolution tells us that BH systems at all multiplicities tend to have lower Eddington ratios~(at fixed $\mathscr{M}$) at lower redshifts, which is due to the general decrease in the AGN luminosity with decreasing redshift at $z\lesssim2-3$. Conversely, this also implies that BH systems of a given Eddington ratio tend to have higher multiplicities at lower redshifts.

\section*{Acknowledgements}
L.B.~acknowledges support from National Science Foundation grant AST-1715413.

\bibliography{references}

\begin{thebibliography}{}
\expandafter\ifx\csname natexlab\endcsname\relax\def\natexlab#1{#1}\fi
\providecommand{\url}[1]{\href{#1}{#1}}
\providecommand{\dodoi}[1]{doi:~\href{http://doi.org/#1}{\nolinkurl{#1}}}
\providecommand{\doeprint}[1]{\href{http://ascl.net/#1}{\nolinkurl{http://ascl.net/#1}}}
\providecommand{\doarXiv}[1]{\href{https://arxiv.org/abs/#1}{\nolinkurl{https://arxiv.org/abs/#1}}}

\bibitem[{{Armitage} \& {Natarajan}(2002)}]{2002ApJ...567L...9A}
{Armitage}, P.~J., \& {Natarajan}, P. 2002, \apjl, 567, L9,
  \dodoi{10.1086/339770}

\bibitem[{{Baker} {et~al.}(2019){Baker}, {Bellovary}, {Bender}, {Berti},
  {Caldwell}, {Camp}, {Conklin}, {Cornish}, {Cutler}, {DeRosa}, {Eracleous},
  {Ferrara}, {Francis}, {Hewitson}, {Holley-Bockelmann}, {Hornschemeier},
  {Hogan}, {Kamai}, {Kelly}, {Shapiro Key}, {Larson}, {Livas},
  {Manthripragada}, {McKenzie}, {McWilliams}, {Mueller}, {Natarajan}, {Numata},
  {Rioux}, {Sankar}, {Schnittman}, {Shoemaker}, {Shoemaker}, {Slutsky},
  {Spero}, {Stebbins}, {Thorpe}, {Vallisneri}, {Ware}, {Wass}, {Yu}, \&
  {Ziemer}}]{2019arXiv190706482B}
{Baker}, J., {Bellovary}, J., {Bender}, P.~L., {et~al.} 2019, arXiv e-prints,
  arXiv:1907.06482.
\newblock \doarXiv{1907.06482}

\bibitem[{{Barnes} \& {Hut}(1986)}]{1986Natur.324..446B}
{Barnes}, J., \& {Hut}, P. 1986, \nat, 324, 446, \dodoi{10.1038/324446a0}

\bibitem[{{Barnes} \& {Hernquist}(1996)}]{1996ApJ...471..115B}
{Barnes}, J.~E., \& {Hernquist}, L. 1996, \apj, 471, 115,
  \dodoi{10.1086/177957}

\bibitem[{{Barret} {et~al.}(2013){Barret}, {Nandra}, {Barcons}, {Fabian}, {den
  Herder}, {Piro}, {Watson}, {Aird}, {Branduardi-Raymont}, {Cappi}, {Carrera},
  {Comastri}, {Costantini}, {Croston}, {Decourchelle}, {Done}, {Dovciak},
  {Ettori}, {Finoguenov}, {Georgakakis}, {Jonker}, {Kaastra}, {Matt}, {Motch},
  {O'Brien}, {Pareschi}, {Pointecouteau}, {Pratt}, {Rauw}, {Reiprich},
  {Sanders}, {Sciortino}, {Willingale}, \& {Wilms}}]{2013sf2a.conf..447B}
{Barret}, D., {Nandra}, K., {Barcons}, X., {et~al.} 2013, in SF2A-2013:
  Proceedings of the Annual meeting of the French Society of Astronomy and
  Astrophysics, ed. L.~{Cambresy}, F.~{Martins}, E.~{Nuss}, \& A.~{Palacios},
  447--453.
\newblock \doarXiv{1310.3814}

\bibitem[{{Begelman} {et~al.}(1980){Begelman}, {Blandford}, \&
  {Rees}}]{1980Natur.287..307B}
{Begelman}, M.~C., {Blandford}, R.~D., \& {Rees}, M.~J. 1980, \nat, 287, 307,
  \dodoi{10.1038/287307a0}

\bibitem[{{Bellovary} {et~al.}(2010){Bellovary}, {Governato}, {Quinn},
  {Wadsley}, {Shen}, \& {Volonteri}}]{2010ApJ...721L.148B}
{Bellovary}, J.~M., {Governato}, F., {Quinn}, T.~R., {et~al.} 2010, \apjl, 721,
  L148, \dodoi{10.1088/2041-8205/721/2/L148}

\bibitem[{{Bhowmick} {et~al.}(2020){Bhowmick}, {Di Matteo}, \&
  {Myers}}]{2020MNRAS.492.5620B}
{Bhowmick}, A.~K., {Di Matteo}, T., \& {Myers}, A.~D. 2020, \mnras, 492, 5620,
  \dodoi{10.1093/mnras/staa172}

\bibitem[{{Bhowmick} {et~al.}(2019){Bhowmick}, {DiMatteo}, {Eftekharzadeh}, \&
  {Myers}}]{2019MNRAS.485.2026B}
{Bhowmick}, A.~K., {DiMatteo}, T., {Eftekharzadeh}, S., \& {Myers}, A.~D. 2019,
  \mnras, 485, 2026, \dodoi{10.1093/mnras/stz519}

\bibitem[{{Blecha} {et~al.}(2013){Blecha}, {Loeb}, \&
  {Narayan}}]{2013MNRAS.429.2594B}
{Blecha}, L., {Loeb}, A., \& {Narayan}, R. 2013, \mnras, 429, 2594,
  \dodoi{10.1093/mnras/sts533}

\bibitem[{{Blecha} {et~al.}(2018){Blecha}, {Snyder}, {Satyapal}, \&
  {Ellison}}]{2018MNRAS.478.3056B}
{Blecha}, L., {Snyder}, G.~F., {Satyapal}, S., \& {Ellison}, S.~L. 2018,
  \mnras, 478, 3056, \dodoi{10.1093/mnras/sty1274}

\bibitem[{{Bonetti} {et~al.}(2016){Bonetti}, {Haardt}, {Sesana}, \&
  {Barausse}}]{2016MNRAS.461.4419B}
{Bonetti}, M., {Haardt}, F., {Sesana}, A., \& {Barausse}, E. 2016, \mnras, 461,
  4419, \dodoi{10.1093/mnras/stw1590}

\bibitem[{{Bonetti} {et~al.}(2019){Bonetti}, {Sesana}, {Haardt}, {Barausse}, \&
  {Colpi}}]{2019MNRAS.486.4044B}
{Bonetti}, M., {Sesana}, A., {Haardt}, F., {Barausse}, E., \& {Colpi}, M. 2019,
  \mnras, 486, 4044, \dodoi{10.1093/mnras/stz903}

\bibitem[{{Bournaud} {et~al.}(2011){Bournaud}, {Dekel}, {Teyssier}, {Cacciato},
  {Daddi}, {Juneau}, \& {Shankar}}]{2011ApJ...741L..33B}
{Bournaud}, F., {Dekel}, A., {Teyssier}, R., {et~al.} 2011, \apjl, 741, L33,
  \dodoi{10.1088/2041-8205/741/2/L33}

\bibitem[{{Capelo} {et~al.}(2015){Capelo}, {Volonteri}, {Dotti}, {Bellovary},
  {Mayer}, \& {Governato}}]{2015MNRAS.447.2123C}
{Capelo}, P.~R., {Volonteri}, M., {Dotti}, M., {et~al.} 2015, \mnras, 447,
  2123, \dodoi{10.1093/mnras/stu2500}

\bibitem[{{Cappelluti} {et~al.}(2009){Cappelluti}, {Brusa}, {Hasinger},
  {Comastri}, {Zamorani}, {Finoguenov}, {Gilli}, {Puccetti}, {Miyaji},
  {Salvato}, {Vignali}, {Aldcroft}, {B{\"o}hringer}, {Brunner}, {Civano},
  {Elvis}, {Fiore}, {Fruscione}, {Griffiths}, {Guzzo}, {Iovino}, {Koekemoer},
  {Mainieri}, {Scoville}, {Shopbell}, {Silverman}, \&
  {Urry}}]{2009A&A...497..635C}
{Cappelluti}, N., {Brusa}, M., {Hasinger}, G., {et~al.} 2009, \aap, 497, 635,
  \dodoi{10.1051/0004-6361/200810794}

\bibitem[{{Chandrasekhar}(1943)}]{1943ApJ....97..255C}
{Chandrasekhar}, S. 1943, \apj, 97, 255, \dodoi{10.1086/144517}

\bibitem[{{Chen} {et~al.}(2009){Chen}, {Wang}, {Yan}, {Hu}, \&
  {Zhang}}]{2009ApJ...695L.130C}
{Chen}, Y.-M., {Wang}, J.-M., {Yan}, C.-S., {Hu}, C., \& {Zhang}, S. 2009,
  \apjl, 695, L130, \dodoi{10.1088/0004-637X/695/2/L130}

\bibitem[{{Cisternas} {et~al.}(2011){Cisternas}, {Jahnke}, {Inskip},
  {Kartaltepe}, {Koekemoer}, {Lisker}, {Robaina}, {Scodeggio}, {Sheth},
  {Trump}, {Andrae}, {Miyaji}, {Lusso}, {Brusa}, {Capak}, {Cappelluti},
  {Civano}, {Ilbert}, {Impey}, {Leauthaud}, {Lilly}, {Salvato}, {Scoville}, \&
  {Taniguchi}}]{2011ApJ...726...57C}
{Cisternas}, M., {Jahnke}, K., {Inskip}, K.~J., {et~al.} 2011, \apj, 726, 57,
  \dodoi{10.1088/0004-637X/726/2/57}

\bibitem[{{Comerford} {et~al.}(2013){Comerford}, {Schluns}, {Greene}, \&
  {Cool}}]{2013ApJ...777...64C}
{Comerford}, J.~M., {Schluns}, K., {Greene}, J.~E., \& {Cool}, R.~J. 2013,
  \apj, 777, 64, \dodoi{10.1088/0004-637X/777/1/64}

\bibitem[{{Davis} {et~al.}(1985){Davis}, {Efstathiou}, {Frenk}, \&
  {White}}]{1985ApJ...292..371D}
{Davis}, M., {Efstathiou}, G., {Frenk}, C.~S., \& {White}, S.~D.~M. 1985, \apj,
  292, 371, \dodoi{10.1086/163168}

\bibitem[{{Desvignes} {et~al.}(2016){Desvignes}, {Caballero}, {Lentati},
  {Verbiest}, {Champion}, {Stappers}, {Janssen}, {Lazarus}, {Os{\l}owski},
  {Babak}, {Bassa}, {Brem}, {Burgay}, {Cognard}, {Gair}, {Graikou},
  {Guillemot}, {Hessels}, {Jessner}, {Jordan}, {Karuppusamy}, {Kramer},
  {Lassus}, {Lazaridis}, {Lee}, {Liu}, {Lyne}, {McKee}, {Mingarelli},
  {Perrodin}, {Petiteau}, {Possenti}, {Purver}, {Rosado}, {Sanidas}, {Sesana},
  {Shaifullah}, {Smits}, {Taylor}, {Theureau}, {Tiburzi}, {van Haasteren}, \&
  {Vecchio}}]{2016MNRAS.458.3341D}
{Desvignes}, G., {Caballero}, R.~N., {Lentati}, L., {et~al.} 2016, \mnras, 458,
  3341, \dodoi{10.1093/mnras/stw483}

\bibitem[{{Di Matteo} {et~al.}(2005){Di Matteo}, {Springel}, \&
  {Hernquist}}]{2005Natur.433..604D}
{Di Matteo}, T., {Springel}, V., \& {Hernquist}, L. 2005, \nat, 433, 604,
  \dodoi{10.1038/nature03335}

\bibitem[{{Eftekharzadeh} {et~al.}(2017){Eftekharzadeh}, {Myers}, {Hennawi},
  {Djorgovski}, {Richards}, {Mahabal}, \& {Graham}}]{2017MNRAS.468...77E}
{Eftekharzadeh}, S., {Myers}, A.~D., {Hennawi}, J.~F., {et~al.} 2017, \mnras,
  468, 77, \dodoi{10.1093/mnras/stx412}

\bibitem[{{Ellison} {et~al.}(2013){Ellison}, {Mendel}, {Patton}, \&
  {Scudder}}]{2013MNRAS.435.3627E}
{Ellison}, S.~L., {Mendel}, J.~T., {Patton}, D.~R., \& {Scudder}, J.~M. 2013,
  \mnras, 435, 3627, \dodoi{10.1093/mnras/stt1562}

\bibitem[{{Ellison} {et~al.}(2011){Ellison}, {Patton}, {Mendel}, \&
  {Scudder}}]{2011MNRAS.418.2043E}
{Ellison}, S.~L., {Patton}, D.~R., {Mendel}, J.~T., \& {Scudder}, J.~M. 2011,
  \mnras, 418, 2043, \dodoi{10.1111/j.1365-2966.2011.19624.x}

\bibitem[{{Ellison} {et~al.}(2019){Ellison}, {Viswanathan}, {Patton},
  {Bottrell}, {McConnachie}, {Gwyn}, \& {Cuillandre}}]{2019MNRAS.487.2491E}
{Ellison}, S.~L., {Viswanathan}, A., {Patton}, D.~R., {et~al.} 2019, \mnras,
  487, 2491, \dodoi{10.1093/mnras/stz1431}

\bibitem[{{Gabor} {et~al.}(2009){Gabor}, {Impey}, {Jahnke}, {Simmons}, {Trump},
  {Koekemoer}, {Brusa}, {Cappelluti}, {Schinnerer}, {Smol{\v{c}}i{\'c}},
  {Salvato}, {Rhodes}, {Mobasher}, {Capak}, {Massey}, {Leauthaud}, \&
  {Scoville}}]{2009ApJ...691..705G}
{Gabor}, J.~M., {Impey}, C.~D., {Jahnke}, K., {et~al.} 2009, \apj, 691, 705,
  \dodoi{10.1088/0004-637X/691/1/705}

\bibitem[{{Gao} {et~al.}(2020){Gao}, {Wang}, {Pearson}, {Gordon}, {Holwerda},
  {Hopkins}, {Brown}, {Bland -Hawthorn}, \& {Owers}}]{2020arXiv200400680G}
{Gao}, F., {Wang}, L., {Pearson}, W.~J., {et~al.} 2020, arXiv e-prints,
  arXiv:2004.00680.
\newblock \doarXiv{2004.00680}

\bibitem[{{Glikman} {et~al.}(2015){Glikman}, {Simmons}, {Mailly}, {Schawinski},
  {Urry}, \& {Lacy}}]{2015ApJ...806..218G}
{Glikman}, E., {Simmons}, B., {Mailly}, M., {et~al.} 2015, \apj, 806, 218,
  \dodoi{10.1088/0004-637X/806/2/218}

\bibitem[{{Goulding} {et~al.}(2018){Goulding}, {Greene}, {Bezanson}, {Greco},
  {Johnson}, {Leauthaud}, {Matsuoka}, {Medezinski}, \&
  {Price-Whelan}}]{2018PASJ...70S..37G}
{Goulding}, A.~D., {Greene}, J.~E., {Bezanson}, R., {et~al.} 2018, \pasj, 70,
  S37, \dodoi{10.1093/pasj/psx135}

\bibitem[{{Gualandris} {et~al.}(2017){Gualandris}, {Read}, {Dehnen}, \&
  {Bortolas}}]{2017MNRAS.464.2301G}
{Gualandris}, A., {Read}, J.~I., {Dehnen}, W., \& {Bortolas}, E. 2017, \mnras,
  464, 2301, \dodoi{10.1093/mnras/stw2528}

\bibitem[{{Harms} {et~al.}(1994){Harms}, {Ford}, {Tsvetanov}, {Hartig},
  {Dressel}, {Kriss}, {Bohlin}, {Davidsen}, {Margon}, \&
  {Kochhar}}]{1994ApJ...435L..35H}
{Harms}, R.~J., {Ford}, H.~C., {Tsvetanov}, Z.~I., {et~al.} 1994, \apjl, 435,
  L35, \dodoi{10.1086/187588}

\bibitem[{{Hennawi} {et~al.}(2006){Hennawi}, {Strauss}, {Oguri}, {Inada},
  {Richards}, {Pindor}, {Schneider}, {Becker}, {Gregg}, {Hall}, {Johnston},
  {Fan}, {Burles}, {Schlegel}, {Gunn}, {Lupton}, {Bahcall}, {Brunner}, \&
  {Brinkmann}}]{2006AJ....131....1H}
{Hennawi}, J.~F., {Strauss}, M.~A., {Oguri}, M., {et~al.} 2006, \aj, 131, 1,
  \dodoi{10.1086/498235}

\bibitem[{{Holley-Bockelmann} \& {Khan}(2015)}]{2015ApJ...810..139H}
{Holley-Bockelmann}, K., \& {Khan}, F.~M. 2015, \apj, 810, 139,
  \dodoi{10.1088/0004-637X/810/2/139}

\bibitem[{{Hopkins} {et~al.}(2006){Hopkins}, {Hernquist}, {Cox}, {Di Matteo},
  {Robertson}, \& {Springel}}]{2006ApJS..163....1H}
{Hopkins}, P.~F., {Hernquist}, L., {Cox}, T.~J., {et~al.} 2006, \apjs, 163, 1,
  \dodoi{10.1086/499298}

\bibitem[{{Hopkins} {et~al.}(2008){Hopkins}, {Hernquist}, {Cox}, \&
  {Kere{\v{s}}}}]{2008ApJS..175..356H}
{Hopkins}, P.~F., {Hernquist}, L., {Cox}, T.~J., \& {Kere{\v{s}}}, D. 2008,
  \apjs, 175, 356, \dodoi{10.1086/524362}

\bibitem[{{Hou} {et~al.}(2020){Hou}, {Li}, \& {Liu}}]{2020arXiv200110686H}
{Hou}, M., {Li}, Z., \& {Liu}, X. 2020, arXiv e-prints, arXiv:2001.10686.
\newblock \doarXiv{2001.10686}

\bibitem[{{Hou} {et~al.}(2019){Hou}, {Liu}, {Guo}, {Li}, {Shen}, \&
  {Green}}]{2019ApJ...882...41H}
{Hou}, M., {Liu}, X., {Guo}, H., {et~al.} 2019, \apj, 882, 41,
  \dodoi{10.3847/1538-4357/ab3225}

\bibitem[{{Kayo} \& {Oguri}(2012)}]{2012MNRAS.424.1363K}
{Kayo}, I., \& {Oguri}, M. 2012, \mnras, 424, 1363,
  \dodoi{10.1111/j.1365-2966.2012.21321.x}

\bibitem[{{Kelley} {et~al.}(2017){Kelley}, {Blecha}, \&
  {Hernquist}}]{2017MNRAS.464.3131K}
{Kelley}, L.~Z., {Blecha}, L., \& {Hernquist}, L. 2017, \mnras, 464, 3131,
  \dodoi{10.1093/mnras/stw2452}

\bibitem[{{Khan} {et~al.}(2013){Khan}, {Holley-Bockelmann}, {Berczik}, \&
  {Just}}]{2013ApJ...773..100K}
{Khan}, F.~M., {Holley-Bockelmann}, K., {Berczik}, P., \& {Just}, A. 2013,
  \apj, 773, 100, \dodoi{10.1088/0004-637X/773/2/100}

\bibitem[{{Kocevski} {et~al.}(2012){Kocevski}, {Faber}, {Mozena}, {Koekemoer},
  {Nandra}, {Rangel}, {Laird}, {Brusa}, {Wuyts}, {Trump}, {Koo}, {Somerville},
  {Bell}, {Lotz}, {Alexander}, {Bournaud}, {Conselice}, {Dahlen}, {Dekel},
  {Donley}, {Dunlop}, {Finoguenov}, {Georgakakis}, {Giavalisco}, {Guo},
  {Grogin}, {Hathi}, {Juneau}, {Kartaltepe}, {Lucas}, {McGrath}, {McIntosh},
  {Mobasher}, {Robaina}, {Rosario}, {Straughn}, {van der Wel}, \&
  {Villforth}}]{2012ApJ...744..148K}
{Kocevski}, D.~D., {Faber}, S.~M., {Mozena}, M., {et~al.} 2012, \apj, 744, 148,
  \dodoi{10.1088/0004-637X/744/2/148}

\bibitem[{{Kocevski} {et~al.}(2015){Kocevski}, {Brightman}, {Nandra},
  {Koekemoer}, {Salvato}, {Aird}, {Bell}, {Hsu}, {Kartaltepe}, {Koo}, {Lotz},
  {McIntosh}, {Mozena}, {Rosario}, \& {Trump}}]{2015ApJ...814..104K}
{Kocevski}, D.~D., {Brightman}, M., {Nandra}, K., {et~al.} 2015, \apj, 814,
  104, \dodoi{10.1088/0004-637X/814/2/104}

\bibitem[{{Kormendy} \& {Kennicutt}(2004)}]{2004ARA&A..42..603K}
{Kormendy}, J., \& {Kennicutt}, Robert~C., J. 2004, \araa, 42, 603,
  \dodoi{10.1146/annurev.astro.42.053102.134024}

\bibitem[{{Kormendy} \& {Richstone}(1992)}]{1992ApJ...393..559K}
{Kormendy}, J., \& {Richstone}, D. 1992, \apj, 393, 559, \dodoi{10.1086/171528}

\bibitem[{{Koss} {et~al.}(2012){Koss}, {Mushotzky}, {Treister}, {Veilleux},
  {Vasudevan}, \& {Trippe}}]{2012ApJ...746L..22K}
{Koss}, M., {Mushotzky}, R., {Treister}, E., {et~al.} 2012, \apjl, 746, L22,
  \dodoi{10.1088/2041-8205/746/2/L22}

\bibitem[{{Koss} {et~al.}(2011){Koss}, {Mushotzky}, {Treister}, {Veilleux},
  {Vasudevan}, {Miller}, {Sand ers}, {Schawinski}, \&
  {Trippe}}]{2011ApJ...735L..42K}
---. 2011, \apjl, 735, L42, \dodoi{10.1088/2041-8205/735/2/L42}

\bibitem[{{Koss} {et~al.}(2018){Koss}, {Blecha}, {Bernhard}, {Hung}, {Lu},
  {Trakhtenbrot}, {Treister}, {Weigel}, {Sartori}, {Mushotzky}, {Schawinski},
  {Ricci}, {Veilleux}, \& {Sanders}}]{2018Natur.563..214K}
{Koss}, M.~J., {Blecha}, L., {Bernhard}, P., {et~al.} 2018, \nat, 563, 214,
  \dodoi{10.1038/s41586-018-0652-7}

\bibitem[{{Kova{\v{c}}evi{\'c}} {et~al.}(2020){Kova{\v{c}}evi{\'c}}, {Yi},
  {Dai}, {Yang}, {{\v{C}}vorovi{\'c}-Hajdinjak}, \&
  {Popovi{\'c}}}]{2020MNRAS.494.4069K}
{Kova{\v{c}}evi{\'c}}, A.~B., {Yi}, T., {Dai}, X., {et~al.} 2020, \mnras, 494,
  4069, \dodoi{10.1093/mnras/staa737}

\bibitem[{{Krumpe} {et~al.}(2018){Krumpe}, {Miyaji}, {Coil}, \&
  {Aceves}}]{2018MNRAS.474.1773K}
{Krumpe}, M., {Miyaji}, T., {Coil}, A.~L., \& {Aceves}, H. 2018, \mnras, 474,
  1773, \dodoi{10.1093/mnras/stx2705}

\bibitem[{{Kumar} \& {Johnson}(2010)}]{2010MNRAS.404.2170K}
{Kumar}, P., \& {Johnson}, J.~L. 2010, \mnras, 404, 2170,
  \dodoi{10.1111/j.1365-2966.2010.16437.x}

\bibitem[{{Lackner} {et~al.}(2014){Lackner}, {Silverman}, {Salvato},
  {Kampczyk}, {Kartaltepe}, {Sanders}, {Capak}, {Civano}, {Halliday}, {Ilbert},
  {Jahnke}, {Koekemoer}, {Lee}, {Le F{\`e}vre}, {Liu}, {Scoville}, {Sheth}, \&
  {Toft}}]{2014AJ....148..137L}
{Lackner}, C.~N., {Silverman}, J.~D., {Salvato}, M., {et~al.} 2014, \aj, 148,
  137, \dodoi{10.1088/0004-6256/148/6/137}

\bibitem[{{LaMassa} {et~al.}(2013){LaMassa}, {Urry}, {Glikman}, {Cappelluti},
  {Civano}, {Comastri}, {Treister}, {B{\"o}hringer}, {Cardamone}, {Chon},
  {Kephart}, {Murray}, {Richards}, {Ross}, {Rozner}, \&
  {Schawinski}}]{2013MNRAS.432.1351L}
{LaMassa}, S.~M., {Urry}, C.~M., {Glikman}, E., {et~al.} 2013, \mnras, 432,
  1351, \dodoi{10.1093/mnras/stt553}

\bibitem[{{Li} {et~al.}(2006){Li}, {Kauffmann}, {Wang}, {White}, {Heckman}, \&
  {Jing}}]{2006MNRAS.373..457L}
{Li}, C., {Kauffmann}, G., {Wang}, L., {et~al.} 2006, \mnras, 373, 457,
  \dodoi{10.1111/j.1365-2966.2006.11079.x}

\bibitem[{{Liu} {et~al.}(2019){Liu}, {Gezari}, {Ayers}, {Burgett}, {Chambers},
  {Hodapp}, {Huber}, {Kudritzki}, {Metcalfe}, {Tonry}, {Wainscoat}, \&
  {Waters}}]{2019ApJ...884...36L}
{Liu}, T., {Gezari}, S., {Ayers}, M., {et~al.} 2019, \apj, 884, 36,
  \dodoi{10.3847/1538-4357/ab40cb}

\bibitem[{{Liu} {et~al.}(2011){Liu}, {Shen}, {Strauss}, \&
  {Hao}}]{2011ApJ...737..101L}
{Liu}, X., {Shen}, Y., {Strauss}, M.~A., \& {Hao}, L. 2011, \apj, 737, 101,
  \dodoi{10.1088/0004-637X/737/2/101}

\bibitem[{{Lusso} {et~al.}(2012){Lusso}, {Comastri}, {Simmons}, {Mignoli},
  {Zamorani}, {Vignali}, {Brusa}, {Shankar}, {Lutz}, {Trump}, {Maiolino},
  {Gilli}, {Bolzonella}, {Puccetti}, {Salvato}, {Impey}, {Civano}, {Elvis},
  {Mainieri}, {Silverman}, {Koekemoer}, {Bongiorno}, {Merloni}, {Berta}, {Le
  Floc'h}, {Magnelli}, {Pozzi}, \& {Riguccini}}]{2012MNRAS.425..623L}
{Lusso}, E., {Comastri}, A., {Simmons}, B.~D., {et~al.} 2012, \mnras, 425, 623,
  \dodoi{10.1111/j.1365-2966.2012.21513.x}

\bibitem[{{Manchester} {et~al.}(2013){Manchester}, {Hobbs}, {Bailes}, {Coles},
  {van Straten}, {Keith}, {Shannon}, {Bhat}, {Brown}, {Burke-Spolaor},
  {Champion}, {Chaudhary}, {Edwards}, {Hampson}, {Hotan}, {Jameson}, {Jenet},
  {Kesteven}, {Khoo}, {Kocz}, {Maciesiak}, {Oslowski}, {Ravi}, {Reynolds},
  {Sarkissian}, {Verbiest}, {Wen}, {Wilson}, {Yardley}, {Yan}, \&
  {You}}]{2013PASA...30...17M}
{Manchester}, R.~N., {Hobbs}, G., {Bailes}, M., {et~al.} 2013, \pasa, 30, e017,
  \dodoi{10.1017/pasa.2012.017}

\bibitem[{{Mannerkoski} {et~al.}(2019){Mannerkoski}, {Johansson}, {Pihajoki},
  {Rantala}, \& {Naab}}]{2019ApJ...887...35M}
{Mannerkoski}, M., {Johansson}, P.~H., {Pihajoki}, P., {Rantala}, A., \&
  {Naab}, T. 2019, \apj, 887, 35, \dodoi{10.3847/1538-4357/ab52f9}

\bibitem[{{Marian} {et~al.}(2019){Marian}, {Jahnke}, {Mechtley}, {Cohen},
  {Husemann}, {Jones}, {Koekemoer}, {Schulze}, {van der Wel}, {Villforth}, \&
  {Windhorst}}]{2019ApJ...882..141M}
{Marian}, V., {Jahnke}, K., {Mechtley}, M., {et~al.} 2019, \apj, 882, 141,
  \dodoi{10.3847/1538-4357/ab385b}

\bibitem[{{Marinacci} {et~al.}(2018){Marinacci}, {Vogelsberger}, {Pakmor},
  {Torrey}, {Springel}, {Hernquist}, {Nelson}, {Weinberger}, {Pillepich},
  {Naiman}, \& {Genel}}]{2018MNRAS.480.5113M}
{Marinacci}, F., {Vogelsberger}, M., {Pakmor}, R., {et~al.} 2018, \mnras, 480,
  5113, \dodoi{10.1093/mnras/sty2206}

\bibitem[{{McAlpine} {et~al.}(2020){McAlpine}, {Harrison}, {Rosario},
  {Alexander}, {Ellison}, {Johansson}, \& {Patton}}]{2020MNRAS.494.5713M}
{McAlpine}, S., {Harrison}, C.~M., {Rosario}, D.~J., {et~al.} 2020, \mnras,
  494, 5713, \dodoi{10.1093/mnras/staa1123}

\bibitem[{{McGreer} {et~al.}(2016){McGreer}, {Eftekharzadeh}, {Myers}, \&
  {Fan}}]{2016AJ....151...61M}
{McGreer}, I.~D., {Eftekharzadeh}, S., {Myers}, A.~D., \& {Fan}, X. 2016, \aj,
  151, 61, \dodoi{10.3847/0004-6256/151/3/61}

\bibitem[{{Milosavljevi{\'c}} \& {Merritt}(2003)}]{2003AIPC..686..201M}
{Milosavljevi{\'c}}, M., \& {Merritt}, D. 2003, in American Institute of
  Physics Conference Series, Vol. 686, The Astrophysics of Gravitational Wave
  Sources, ed. J.~M. {Centrella}, 201--210, \dodoi{10.1063/1.1629432}

\bibitem[{{Miyoshi} {et~al.}(1995){Miyoshi}, {Moran}, {Herrnstein},
  {Greenhill}, {Nakai}, {Diamond}, \& {Inoue}}]{1995Natur.373..127M}
{Miyoshi}, M., {Moran}, J., {Herrnstein}, J., {et~al.} 1995, \nat, 373, 127,
  \dodoi{10.1038/373127a0}

\bibitem[{{Mushotzky}(2018)}]{2018SPIE10699E..29M}
{Mushotzky}, R. 2018, in Society of Photo-Optical Instrumentation Engineers
  (SPIE) Conference Series, Vol. 10699, \procspie, 1069929,
  \dodoi{10.1117/12.2310003}

\bibitem[{{Nandra} {et~al.}(2013){Nandra}, {Barret}, {Barcons}, {Fabian}, {den
  Herder}, {Piro}, {Watson}, {Adami}, {Aird}, {Afonso}, {Alexander},
  {Argiroffi}, {Amati}, {Arnaud}, {Atteia}, {Audard}, {Badenes}, {Ballet},
  {Ballo}, {Bamba}, {Bhardwaj}, {Stefano Battistelli}, {Becker}, {De Becker},
  {Behar}, {Bianchi}, {Biffi}, {B{\^\i}rzan}, {Bocchino}, {Bogdanov}, {Boirin},
  {Boller}, {Borgani}, {Borm}, {Bouch{\'e}}, {Bourdin}, {Bower}, {Braito},
  {Branchini}, {Branduardi-Raymont}, {Bregman}, {Brenneman}, {Brightman},
  {Br{\"u}ggen}, {Buchner}, {Bulbul}, {Brusa}, {Bursa}, {Caccianiga},
  {Cackett}, {Campana}, {Cappelluti}, {Cappi}, {Carrera}, {Ceballos},
  {Christensen}, {Chu}, {Churazov}, {Clerc}, {Corbel}, {Corral}, {Comastri},
  {Costantini}, {Croston}, {Dadina}, {D'Ai}, {Decourchelle}, {Della Ceca},
  {Dennerl}, {Dolag}, {Done}, {Dovciak}, {Drake}, {Eckert}, {Edge}, {Ettori},
  {Ezoe}, {Feigelson}, {Fender}, {Feruglio}, {Finoguenov}, {Fiore}, {Galeazzi},
  {Gallagher}, {Gandhi}, {Gaspari}, {Gastaldello}, {Georgakakis},
  {Georgantopoulos}, {Gilfanov}, {Gitti}, {Gladstone}, {Goosmann}, {Gosset},
  {Grosso}, {Guedel}, {Guerrero}, {Haberl}, {Hardcastle}, {Heinz}, {Alonso
  Herrero}, {Herv{\'e}}, {Holmstrom}, {Iwasawa}, {Jonker}, {Kaastra}, {Kara},
  {Karas}, {Kastner}, {King}, {Kosenko}, {Koutroumpa}, {Kraft}, {Kreykenbohm},
  {Lallement}, {Lanzuisi}, {Lee}, {Lemoine-Goumard}, {Lobban}, {Lodato},
  {Lovisari}, {Lotti}, {McCharthy}, {McNamara}, {Maggio}, {Maiolino}, {De
  Marco}, {de Martino}, {Mateos}, {Matt}, {Maughan}, {Mazzotta}, {Mendez},
  {Merloni}, {Micela}, {Miceli}, {Mignani}, {Miller}, {Miniutti}, {Molendi},
  {Montez}, {Moretti}, {Motch}, {Naz{\'e}}, {Nevalainen}, {Nicastro}, {Nulsen},
  {Ohashi}, {O'Brien}, {Osborne}, {Oskinova}, {Pacaud}, {Paerels}, {Page},
  {Papadakis}, {Pareschi}, {Petre}, {Petrucci}, {Piconcelli}, {Pillitteri},
  {Pinto}, {de Plaa}, {Pointecouteau}, {Ponman}, {Ponti}, {Porquet}, {Pounds},
  {Pratt}, {Predehl}, {Proga}, {Psaltis}, {Rafferty}, {Ramos-Ceja}, {Ranalli},
  {Rasia}, {Rau}, {Rauw}, {Rea}, {Read}, {Reeves}, {Reiprich}, {Renaud},
  {Reynolds}, {Risaliti}, {Rodriguez}, {Rodriguez Hidalgo}, {Roncarelli},
  {Rosario}, {Rossetti}, {Rozanska}, {Rovilos}, {Salvaterra}, {Salvato}, {Di
  Salvo}, {Sanders}, {Sanz-Forcada}, {Schawinski}, {Schaye}, {Schwope},
  {Sciortino}, {Severgnini}, {Shankar}, {Sijacki}, {Sim}, {Schmid}, {Smith},
  {Steiner}, {Stelzer}, {Stewart}, {Strohmayer}, {Str{\"u}der}, {Sun}, {Takei},
  {Tatischeff}, {Tiengo}, {Tombesi}, {Trinchieri}, {Tsuru}, {Ud-Doula},
  {Ursino}, {Valencic}, {Vanzella}, {Vaughan}, {Vignali}, {Vink}, {Vito},
  {Volonteri}, {Wang}, {Webb}, {Willingale}, {Wilms}, {Wise}, {Worrall},
  {Young}, {Zampieri}, {In't Zand}, {Zane}, {Zezas}, {Zhang}, \&
  {Zhuravleva}}]{2013arXiv1306.2307N}
{Nandra}, K., {Barret}, D., {Barcons}, X., {et~al.} 2013, arXiv e-prints,
  arXiv:1306.2307.
\newblock \doarXiv{1306.2307}

\bibitem[{{Nasim} {et~al.}(2020){Nasim}, {Gualandris}, {Read}, {Dehnen},
  {Delorme}, \& {Antonini}}]{2020arXiv200414399N}
{Nasim}, I., {Gualandris}, A., {Read}, J., {et~al.} 2020, arXiv e-prints,
  arXiv:2004.14399.
\newblock \doarXiv{2004.14399}

\bibitem[{{Nelson} {et~al.}(2015){Nelson}, {Pillepich}, {Genel},
  {Vogelsberger}, {Springel}, {Torrey}, {Rodriguez-Gomez}, {Sijacki}, {Snyder},
  {Griffen}, {Marinacci}, {Blecha}, {Sales}, {Xu}, \&
  {Hernquist}}]{2015A&C....13...12N}
{Nelson}, D., {Pillepich}, A., {Genel}, S., {et~al.} 2015, Astronomy and
  Computing, 13, 12, \dodoi{10.1016/j.ascom.2015.09.003}

\bibitem[{{Nelson} {et~al.}(2018){Nelson}, {Pillepich}, {Springel},
  {Weinberger}, {Hernquist}, {Pakmor}, {Genel}, {Torrey}, {Vogelsberger},
  {Kauffmann}, {Marinacci}, \& {Naiman}}]{2018MNRAS.475..624N}
{Nelson}, D., {Pillepich}, A., {Springel}, V., {et~al.} 2018, \mnras, 475, 624,
  \dodoi{10.1093/mnras/stx3040}

\bibitem[{{Nelson} {et~al.}(2019){Nelson}, {Pillepich}, {Springel}, {Pakmor},
  {Weinberger}, {Genel}, {Torrey}, {Vogelsberger}, {Marinacci}, \&
  {Hernquist}}]{2019MNRAS.490.3234N}
---. 2019, \mnras, 490, 3234, \dodoi{10.1093/mnras/stz2306}

\bibitem[{{Ogiya} {et~al.}(2020){Ogiya}, {Hahn}, {Mingarelli}, \&
  {Volonteri}}]{2020MNRAS.493.3676O}
{Ogiya}, G., {Hahn}, O., {Mingarelli}, C. M.~F., \& {Volonteri}, M. 2020,
  \mnras, 493, 3676, \dodoi{10.1093/mnras/staa444}

\bibitem[{{Oh} {et~al.}(2018){Oh}, {Koss}, {Markwardt}, {Schawinski},
  {Baumgartner}, {Barthelmy}, {Cenko}, {Gehrels}, {Mushotzky}, {Petulante},
  {Ricci}, {Lien}, \& {Trakhtenbrot}}]{2018ApJS..235....4O}
{Oh}, K., {Koss}, M., {Markwardt}, C.~B., {et~al.} 2018, \apjs, 235, 4,
  \dodoi{10.3847/1538-4365/aaa7fd}

\bibitem[{{Pakmor} {et~al.}(2011){Pakmor}, {Bauer}, \&
  {Springel}}]{2011MNRAS.418.1392P}
{Pakmor}, R., {Bauer}, A., \& {Springel}, V. 2011, \mnras, 418, 1392,
  \dodoi{10.1111/j.1365-2966.2011.19591.x}

\bibitem[{{Pakmor} {et~al.}(2016){Pakmor}, {Pfrommer}, {Simpson}, {Kannan}, \&
  {Springel}}]{2016MNRAS.462.2603P}
{Pakmor}, R., {Pfrommer}, C., {Simpson}, C.~M., {Kannan}, R., \& {Springel}, V.
  2016, \mnras, 462, 2603, \dodoi{10.1093/mnras/stw1761}

\bibitem[{Perets \& Alexander(2008)}]{Perets_2008}
Perets, H.~B., \& Alexander, T. 2008, The Astrophysical Journal, 677, 146,
  \dodoi{10.1086/527525}

\bibitem[{{Pfeifle} {et~al.}(2019){Pfeifle}, {Satyapal}, {Manzano-King},
  {Cann}, {Sexton}, {Rothberg}, {Canalizo}, {Ricci}, {Blecha}, {Ellison},
  {Gliozzi}, {Secrest}, {Constantin}, \& {Harvey}}]{2019ApJ...883..167P}
{Pfeifle}, R.~W., {Satyapal}, S., {Manzano-King}, C., {et~al.} 2019, \apj, 883,
  167, \dodoi{10.3847/1538-4357/ab3a9b}

\bibitem[{Pillepich {et~al.}(2017)Pillepich, Springel, Nelson, Genel, Naiman,
  Pakmor, Hernquist, Torrey, Vogelsberger, Weinberger, \&
  Marinacci}]{10.1093/mnras/stx2656}
Pillepich, A., Springel, V., Nelson, D., {et~al.} 2017, Monthly Notices of the
  Royal Astronomical Society, 473, 4077, \dodoi{10.1093/mnras/stx2656}

\bibitem[{{Pillepich} {et~al.}(2018{\natexlab{a}}){Pillepich}, {Springel},
  {Nelson}, {Genel}, {Naiman}, {Pakmor}, {Hernquist}, {Torrey}, {Vogelsberger},
  {Weinberger}, \& {Marinacci}}]{2018MNRAS.473.4077P}
{Pillepich}, A., {Springel}, V., {Nelson}, D., {et~al.} 2018{\natexlab{a}},
  \mnras, 473, 4077, \dodoi{10.1093/mnras/stx2656}

\bibitem[{{Pillepich} {et~al.}(2018{\natexlab{b}}){Pillepich}, {Nelson},
  {Hernquist}, {Springel}, {Pakmor}, {Torrey}, {Weinberger}, {Genel}, {Naiman},
  {Marinacci}, \& {Vogelsberger}}]{2018MNRAS.475..648P}
{Pillepich}, A., {Nelson}, D., {Hernquist}, L., {et~al.} 2018{\natexlab{b}},
  \mnras, 475, 648, \dodoi{10.1093/mnras/stx3112}

\bibitem[{{Pillepich} {et~al.}(2019){Pillepich}, {Nelson}, {Springel},
  {Pakmor}, {Torrey}, {Weinberger}, {Vogelsberger}, {Marinacci}, {Genel}, {van
  der Wel}, \& {Hernquist}}]{2019MNRAS.490.3196P}
{Pillepich}, A., {Nelson}, D., {Springel}, V., {et~al.} 2019, \mnras, 490,
  3196, \dodoi{10.1093/mnras/stz2338}

\bibitem[{{Planck Collaboration} {et~al.}(2016){Planck Collaboration}, {Ade},
  {Aghanim}, {Arnaud}, {Ashdown}, {Aumont}, {Baccigalupi}, {Banday},
  {Barreiro}, {Bartlett}, {Bartolo}, {Battaner}, {Battye}, {Benabed},
  {Beno{\^\i}t}, {Benoit-L{\'e}vy}, {Bernard}, {Bersanelli}, {Bielewicz},
  {Bock}, {Bonaldi}, {Bonavera}, {Bond}, {Borrill}, {Bouchet}, {Boulanger},
  {Bucher}, {Burigana}, {Butler}, {Calabrese}, {Cardoso}, {Catalano},
  {Challinor}, {Chamballu}, {Chary}, {Chiang}, {Chluba}, {Christensen},
  {Church}, {Clements}, {Colombi}, {Colombo}, {Combet}, {Coulais}, {Crill},
  {Curto}, {Cuttaia}, {Danese}, {Davies}, {Davis}, {de Bernardis}, {de Rosa},
  {de Zotti}, {Delabrouille}, {D{\'e}sert}, {Di Valentino}, {Dickinson},
  {Diego}, {Dolag}, {Dole}, {Donzelli}, {Dor{\'e}}, {Douspis}, {Ducout},
  {Dunkley}, {Dupac}, {Efstathiou}, {Elsner}, {En{\ss}lin}, {Eriksen},
  {Farhang}, {Fergusson}, {Finelli}, {Forni}, {Frailis}, {Fraisse},
  {Franceschi}, {Frejsel}, {Galeotta}, {Galli}, {Ganga}, {Gauthier}, {Gerbino},
  {Ghosh}, {Giard}, {Giraud-H{\'e}raud}, {Giusarma}, {Gjerl{\o}w},
  {Gonz{\'a}lez-Nuevo}, {G{\'o}rski}, {Gratton}, {Gregorio}, {Gruppuso},
  {Gudmundsson}, {Hamann}, {Hansen}, {Hanson}, {Harrison}, {Helou},
  {Henrot-Versill{\'e}}, {Hern{\'a}ndez-Monteagudo}, {Herranz}, {Hildebrand t},
  {Hivon}, {Hobson}, {Holmes}, {Hornstrup}, {Hovest}, {Huang}, {Huffenberger},
  {Hurier}, {Jaffe}, {Jaffe}, {Jones}, {Juvela}, {Keih{\"a}nen}, {Keskitalo},
  {Kisner}, {Kneissl}, {Knoche}, {Knox}, {Kunz}, {Kurki-Suonio}, {Lagache},
  {L{\"a}hteenm{\"a}ki}, {Lamarre}, {Lasenby}, {Lattanzi}, {Lawrence}, {Leahy},
  {Leonardi}, {Lesgourgues}, {Levrier}, {Lewis}, {Liguori}, {Lilje},
  {Linden-V{\o}rnle}, {L{\'o}pez-Caniego}, {Lubin}, {Mac{\'\i}as-P{\'e}rez},
  {Maggio}, {Maino}, {Mandolesi}, {Mangilli}, {Marchini}, {Maris}, {Martin},
  {Martinelli}, {Mart{\'\i}nez-Gonz{\'a}lez}, {Masi}, {Matarrese}, {McGehee},
  {Meinhold}, {Melchiorri}, {Melin}, {Mendes}, {Mennella}, {Migliaccio},
  {Millea}, {Mitra}, {Miville-Desch{\^e}nes}, {Moneti}, {Montier}, {Morgante},
  {Mortlock}, {Moss}, {Munshi}, {Murphy}, {Naselsky}, {Nati}, {Natoli},
  {Netterfield}, {N{\o}rgaard-Nielsen}, {Noviello}, {Novikov}, {Novikov},
  {Oxborrow}, {Paci}, {Pagano}, {Pajot}, {Paladini}, {Paoletti}, {Partridge},
  {Pasian}, {Patanchon}, {Pearson}, {Perdereau}, {Perotto}, {Perrotta},
  {Pettorino}, {Piacentini}, {Piat}, {Pierpaoli}, {Pietrobon}, {Plaszczynski},
  {Pointecouteau}, {Polenta}, {Popa}, {Pratt}, {Pr{\'e}zeau}, {Prunet},
  {Puget}, {Rachen}, {Reach}, {Rebolo}, {Reinecke}, {Remazeilles}, {Renault},
  {Renzi}, {Ristorcelli}, {Rocha}, {Rosset}, {Rossetti}, {Roudier},
  {Rouill{\'e} d'Orfeuil}, {Rowan-Robinson}, {Rubi{\~n}o-Mart{\'\i}n},
  {Rusholme}, {Said}, {Salvatelli}, {Salvati}, {Sandri}, {Santos},
  {Savelainen}, {Savini}, {Scott}, {Seiffert}, {Serra}, {Shellard}, {Spencer},
  {Spinelli}, {Stolyarov}, {Stompor}, {Sudiwala}, {Sunyaev}, {Sutton},
  {Suur-Uski}, {Sygnet}, {Tauber}, {Terenzi}, {Toffolatti}, {Tomasi},
  {Tristram}, {Trombetti}, {Tucci}, {Tuovinen}, {T{\"u}rler}, {Umana},
  {Valenziano}, {Valiviita}, {Van Tent}, {Vielva}, {Villa}, {Wade}, {Wandelt},
  {Wehus}, {White}, {White}, {Wilkinson}, {Yvon}, {Zacchei}, \&
  {Zonca}}]{2016A&A...594A..13P}
{Planck Collaboration}, {Ade}, P.~A.~R., {Aghanim}, N., {et~al.} 2016, \aap,
  594, A13, \dodoi{10.1051/0004-6361/201525830}

\bibitem[{{Powell} {et~al.}(2020){Powell}, {Urry}, {Cappelluti}, {Johnson},
  {LaMassa}, {Ananna}, \& {Kollmann}}]{2020ApJ...891...41P}
{Powell}, M.~C., {Urry}, C.~M., {Cappelluti}, N., {et~al.} 2020, \apj, 891, 41,
  \dodoi{10.3847/1538-4357/ab6e65}

\bibitem[{{Qu} {et~al.}(2017){Qu}, {Helly}, {Bower}, {Theuns}, {Crain},
  {Frenk}, {Furlong}, {McAlpine}, {Schaller}, {Schaye}, \&
  {White}}]{2017MNRAS.464.1659Q}
{Qu}, Y., {Helly}, J.~C., {Bower}, R.~G., {et~al.} 2017, \mnras, 464, 1659,
  \dodoi{10.1093/mnras/stw2437}

\bibitem[{{Quinlan}(1996)}]{1996NewA....1...35Q}
{Quinlan}, G.~D. 1996, \na, 1, 35, \dodoi{10.1016/S1384-1076(96)00003-6}

\bibitem[{{Rafikov}(2016)}]{2016ApJ...827..111R}
{Rafikov}, R.~R. 2016, \apj, 827, 111, \dodoi{10.3847/0004-637X/827/2/111}

\bibitem[{{Ransom} {et~al.}(2019){Ransom}, {Brazier}, {Chatterjee}, {Cohen},
  {Cordes}, {DeCesar}, {Demorest}, {Hazboun}, {Lam}, {Lynch}, {McLaughlin},
  {Ransom}, {Siemens}, {Taylor}, \& {Vigeland}}]{2019BAAS...51g.195R}
{Ransom}, S., {Brazier}, A., {Chatterjee}, S., {et~al.} 2019, in \baas,
  Vol.~51, 195.
\newblock \doarXiv{1908.05356}

\bibitem[{{Ravi} {et~al.}(2014){Ravi}, {Wyithe}, {Shannon}, {Hobbs}, \&
  {Manchester}}]{2014MNRAS.442...56R}
{Ravi}, V., {Wyithe}, J.~S.~B., {Shannon}, R.~M., {Hobbs}, G., \& {Manchester},
  R.~N. 2014, \mnras, 442, 56, \dodoi{10.1093/mnras/stu779}

\bibitem[{{Ricci} {et~al.}(2017){Ricci}, {Bauer}, {Treister}, {Schawinski},
  {Privon}, {Blecha}, {Arevalo}, {Armus}, {Harrison}, {Ho}, {Iwasawa},
  {Sanders}, \& {Stern}}]{2017MNRAS.468.1273R}
{Ricci}, C., {Bauer}, F.~E., {Treister}, E., {et~al.} 2017, \mnras, 468, 1273,
  \dodoi{10.1093/mnras/stx173}

\bibitem[{{Rodriguez} {et~al.}(2006){Rodriguez}, {Taylor}, {Zavala}, {Peck},
  {Pollack}, \& {Romani}}]{2006ApJ...646...49R}
{Rodriguez}, C., {Taylor}, G.~B., {Zavala}, R.~T., {et~al.} 2006, \apj, 646,
  49, \dodoi{10.1086/504825}

\bibitem[{{Rodriguez-Gomez} {et~al.}(2016){Rodriguez-Gomez}, {Pillepich},
  {Sales}, {Genel}, {Vogelsberger}, {Zhu}, {Wellons}, {Nelson}, {Torrey},
  {Springel}, {Ma}, \& {Hernquist}}]{2016MNRAS.458.2371R}
{Rodriguez-Gomez}, V., {Pillepich}, A., {Sales}, L.~V., {et~al.} 2016, \mnras,
  458, 2371, \dodoi{10.1093/mnras/stw456}

\bibitem[{{Ryu} {et~al.}(2018){Ryu}, {Perna}, {Haiman}, {Ostriker}, \&
  {Stone}}]{2018MNRAS.473.3410R}
{Ryu}, T., {Perna}, R., {Haiman}, Z., {Ostriker}, J.~P., \& {Stone}, N.~C.
  2018, \mnras, 473, 3410, \dodoi{10.1093/mnras/stx2524}

\bibitem[{{S{\'a}nchez-Salcedo} \& {Chametla}(2014)}]{2014ApJ...794..167S}
{S{\'a}nchez-Salcedo}, F.~J., \& {Chametla}, R.~O. 2014, \apj, 794, 167,
  \dodoi{10.1088/0004-637X/794/2/167}

\bibitem[{{Sanders} \& {Mirabel}(1996)}]{1996ARA&A..34..749S}
{Sanders}, D.~B., \& {Mirabel}, I.~F. 1996, \araa, 34, 749,
  \dodoi{10.1146/annurev.astro.34.1.749}

\bibitem[{{Sanders} {et~al.}(1988){Sanders}, {Soifer}, {Elias}, {Madore},
  {Matthews}, {Neugebauer}, \& {Scoville}}]{1988ApJ...325...74S}
{Sanders}, D.~B., {Soifer}, B.~T., {Elias}, J.~H., {et~al.} 1988, \apj, 325,
  74, \dodoi{10.1086/165983}

\bibitem[{{Satyapal} {et~al.}(2014){Satyapal}, {Ellison}, {McAlpine}, {Hickox},
  {Patton}, \& {Mendel}}]{2014MNRAS.441.1297S}
{Satyapal}, S., {Ellison}, S.~L., {McAlpine}, W., {et~al.} 2014, \mnras, 441,
  1297, \dodoi{10.1093/mnras/stu650}

\bibitem[{{Schawinski} {et~al.}(2012){Schawinski}, {Simmons}, {Urry},
  {Treister}, \& {Glikman}}]{2012MNRAS.425L..61S}
{Schawinski}, K., {Simmons}, B.~D., {Urry}, C.~M., {Treister}, E., \&
  {Glikman}, E. 2012, \mnras, 425, L61,
  \dodoi{10.1111/j.1745-3933.2012.01302.x}

\bibitem[{{Schaye} {et~al.}(2015){Schaye}, {Crain}, {Bower}, {Furlong},
  {Schaller}, {Theuns}, {Dalla Vecchia}, {Frenk}, {McCarthy}, {Helly},
  {Jenkins}, {Rosas-Guevara}, {White}, {Baes}, {Booth}, {Camps}, {Navarro},
  {Qu}, {Rahmati}, {Sawala}, {Thomas}, \& {Trayford}}]{2015MNRAS.446..521S}
{Schaye}, J., {Crain}, R.~A., {Bower}, R.~G., {et~al.} 2015, \mnras, 446, 521,
  \dodoi{10.1093/mnras/stu2058}

\bibitem[{{Schneider} {et~al.}(2000){Schneider}, {Fan}, {Strauss}, {Gunn},
  {Richards}, {Knapp}, {Lupton}, {Saxe}, {Anderson}, {Bahcall}, {Brinkmann},
  {Brunner}, {Csabai}, {Fukugita}, {Hennessy}, {Hindsley}, {Ivezi{\'c}},
  {Nichol}, {Pier}, \& {York}}]{2000AJ....120.2183S}
{Schneider}, D.~P., {Fan}, X., {Strauss}, M.~A., {et~al.} 2000, \aj, 120, 2183,
  \dodoi{10.1086/316834}

\bibitem[{{Secrest} {et~al.}(2020){Secrest}, {Ellison}, {Satyapal}, \&
  {Blecha}}]{2020arXiv200601850S}
{Secrest}, N., {Ellison}, S., {Satyapal}, S., \& {Blecha}, L. 2020, arXiv
  e-prints, arXiv:2006.01850.
\newblock \doarXiv{2006.01850}

\bibitem[{{Sesana}(2010)}]{2010ApJ...719..851S}
{Sesana}, A. 2010, \apj, 719, 851, \dodoi{10.1088/0004-637X/719/1/851}

\bibitem[{{Silverman} {et~al.}(2011){Silverman}, {Kampczyk}, {Jahnke},
  {Andrae}, {Lilly}, {Elvis}, {Civano}, {Mainieri}, {Vignali}, {Zamorani},
  {Nair}, {Le F{\`e}vre}, {de Ravel}, {Bardelli}, {Bongiorno}, {Bolzonella},
  {Cappi}, {Caputi}, {Carollo}, {Contini}, {Coppa}, {Cucciati}, {de la Torre},
  {Franzetti}, {Garilli}, {Halliday}, {Hasinger}, {Iovino}, {Knobel},
  {Koekemoer}, {Kova{\v{c}}}, {Lamareille}, {Le Borgne}, {Le Brun}, {Maier},
  {Mignoli}, {Pello}, {P{\'e}rez-Montero}, {Ricciardelli}, {Peng}, {Scodeggio},
  {Tanaka}, {Tasca}, {Tresse}, {Vergani}, {Zucca}, {Brusa}, {Cappelluti},
  {Comastri}, {Finoguenov}, {Fu}, {Gilli}, {Hao}, {Ho}, \&
  {Salvato}}]{2011ApJ...743....2S}
{Silverman}, J.~D., {Kampczyk}, P., {Jahnke}, K., {et~al.} 2011, \apj, 743, 2,
  \dodoi{10.1088/0004-637X/743/1/2}

\bibitem[{{Snyder} {et~al.}(2013){Snyder}, {Hayward}, {Sajina}, {Jonsson},
  {Cox}, {Hernquist}, {Hopkins}, \& {Yan}}]{2013ApJ...768..168S}
{Snyder}, G.~F., {Hayward}, C.~C., {Sajina}, A., {et~al.} 2013, \apj, 768, 168,
  \dodoi{10.1088/0004-637X/768/2/168}

\bibitem[{{Springel}(2010)}]{2010MNRAS.401..791S}
{Springel}, V. 2010, \mnras, 401, 791, \dodoi{10.1111/j.1365-2966.2009.15715.x}

\bibitem[{{Springel} \& {Hernquist}(2003)}]{2003MNRAS.339..289S}
{Springel}, V., \& {Hernquist}, L. 2003, \mnras, 339, 289,
  \dodoi{10.1046/j.1365-8711.2003.06206.x}

\bibitem[{{Springel} {et~al.}(2001){Springel}, {Yoshida}, \&
  {White}}]{2001NewA....6...79S}
{Springel}, V., {Yoshida}, N., \& {White}, S. D.~M. 2001, \na, 6, 79,
  \dodoi{10.1016/S1384-1076(01)00042-2}

\bibitem[{{Springel} {et~al.}(2018){Springel}, {Pakmor}, {Pillepich},
  {Weinberger}, {Nelson}, {Hernquist}, {Vogelsberger}, {Genel}, {Torrey},
  {Marinacci}, \& {Naiman}}]{2018MNRAS.475..676S}
{Springel}, V., {Pakmor}, R., {Pillepich}, A., {et~al.} 2018, \mnras, 475, 676,
  \dodoi{10.1093/mnras/stx3304}

\bibitem[{{The Lynx Team}(2018)}]{2018arXiv180909642T}
{The Lynx Team}. 2018, arXiv e-prints, arXiv:1809.09642.
\newblock \doarXiv{1809.09642}

\bibitem[{{Urrutia} {et~al.}(2008){Urrutia}, {Lacy}, \&
  {Becker}}]{2008ApJ...674...80U}
{Urrutia}, T., {Lacy}, M., \& {Becker}, R.~H. 2008, \apj, 674, 80,
  \dodoi{10.1086/523959}

\bibitem[{{Vasiliev} {et~al.}(2015){Vasiliev}, {Antonini}, \&
  {Merritt}}]{2015ApJ...810...49V}
{Vasiliev}, E., {Antonini}, F., \& {Merritt}, D. 2015, \apj, 810, 49,
  \dodoi{10.1088/0004-637X/810/1/49}

\bibitem[{{Vasudevan} {et~al.}(2010){Vasudevan}, {Fabian}, {Gandhi}, {Winter},
  \& {Mushotzky}}]{2010MNRAS.402.1081V}
{Vasudevan}, R.~V., {Fabian}, A.~C., {Gandhi}, P., {Winter}, L.~M., \&
  {Mushotzky}, R.~F. 2010, \mnras, 402, 1081,
  \dodoi{10.1111/j.1365-2966.2009.15936.x}

\bibitem[{{Veilleux} {et~al.}(2009){Veilleux}, {Rupke}, {Kim}, {Genzel},
  {Sturm}, {Lutz}, {Contursi}, {Schweitzer}, {Tacconi}, {Netzer}, {Sternberg},
  {Mihos}, {Baker}, {Mazzarella}, {Lord}, {Sanders}, {Stockton}, {Joseph}, \&
  {Barnes}}]{2009ApJS..182..628V}
{Veilleux}, S., {Rupke}, D.~S.~N., {Kim}, D.~C., {et~al.} 2009, \apjs, 182,
  628, \dodoi{10.1088/0067-0049/182/2/628}

\bibitem[{{Verbiest} {et~al.}(2016){Verbiest}, {Lentati}, {Hobbs}, {van
  Haasteren}, {Demorest}, {Janssen}, {Wang}, {Desvignes}, {Caballero}, {Keith},
  {Champion}, {Arzoumanian}, {Babak}, {Bassa}, {Bhat}, {Brazier}, {Brem},
  {Burgay}, {Burke-Spolaor}, {Chamberlin}, {Chatterjee}, {Christy}, {Cognard},
  {Cordes}, {Dai}, {Dolch}, {Ellis}, {Ferdman}, {Fonseca}, {Gair},
  {Garver-Daniels}, {Gentile}, {Gonzalez}, {Graikou}, {Guillemot}, {Hessels},
  {Jones}, {Karuppusamy}, {Kerr}, {Kramer}, {Lam}, {Lasky}, {Lassus},
  {Lazarus}, {Lazio}, {Lee}, {Levin}, {Liu}, {Lynch}, {Lyne}, {Mckee},
  {McLaughlin}, {McWilliams}, {Madison}, {Manchester}, {Mingarelli}, {Nice},
  {Os{\l}owski}, {Palliyaguru}, {Pennucci}, {Perera}, {Perrodin}, {Possenti},
  {Petiteau}, {Ransom}, {Reardon}, {Rosado}, {Sanidas}, {Sesana}, {Shaifullah},
  {Shannon}, {Siemens}, {Simon}, {Smits}, {Spiewak}, {Stairs}, {Stappers},
  {Stinebring}, {Stovall}, {Swiggum}, {Taylor}, {Theureau}, {Tiburzi},
  {Toomey}, {Vallisneri}, {van Straten}, {Vecchio}, {Wang}, {Wen}, {You},
  {Zhu}, \& {Zhu}}]{2016MNRAS.458.1267V}
{Verbiest}, J.~P.~W., {Lentati}, L., {Hobbs}, G., {et~al.} 2016, \mnras, 458,
  1267, \dodoi{10.1093/mnras/stw347}

\bibitem[{{Villforth} {et~al.}(2014){Villforth}, {Hamann}, {Rosario},
  {Santini}, {McGrath}, {van der Wel}, {Chang}, {Guo}, {Dahlen}, {Bell},
  {Conselice}, {Croton}, {Dekel}, {Faber}, {Grogin}, {Hamilton}, {Hopkins},
  {Juneau}, {Kartaltepe}, {Kocevski}, {Koekemoer}, {Koo}, {Lotz}, {McIntosh},
  {Mozena}, {Somerville}, \& {Wild}}]{2014MNRAS.439.3342V}
{Villforth}, C., {Hamann}, F., {Rosario}, D.~J., {et~al.} 2014, \mnras, 439,
  3342, \dodoi{10.1093/mnras/stu173}

\bibitem[{{Villforth} {et~al.}(2017){Villforth}, {Hamilton}, {Pawlik},
  {Hewlett}, {Rowlands}, {Herbst}, {Shankar}, {Fontana}, {Hamann}, {Koekemoer},
  {Pforr}, {Trump}, \& {Wuyts}}]{2017MNRAS.466..812V}
{Villforth}, C., {Hamilton}, T., {Pawlik}, M.~M., {et~al.} 2017, \mnras, 466,
  812, \dodoi{10.1093/mnras/stw3037}

\bibitem[{{Vogelsberger} {et~al.}(2013){Vogelsberger}, {Genel}, {Sijacki},
  {Torrey}, {Springel}, \& {Hernquist}}]{2013MNRAS.436.3031V}
{Vogelsberger}, M., {Genel}, S., {Sijacki}, D., {et~al.} 2013, \mnras, 436,
  3031, \dodoi{10.1093/mnras/stt1789}

\bibitem[{{Vogelsberger} {et~al.}(2014){Vogelsberger}, {Genel}, {Springel},
  {Torrey}, {Sijacki}, {Xu}, {Snyder}, {Nelson}, \&
  {Hernquist}}]{2014MNRAS.444.1518V}
{Vogelsberger}, M., {Genel}, S., {Springel}, V., {et~al.} 2014, \mnras, 444,
  1518, \dodoi{10.1093/mnras/stu1536}

\bibitem[{{Vogelsberger} {et~al.}(2019){Vogelsberger}, {Nelson}, {Pillepich},
  {Shen}, {Marinacci}, {Springel}, {Pakmor}, {Tacchella}, {Weinberger},
  {Torrey}, \& {Hernquist}}]{2019arXiv190407238V}
{Vogelsberger}, M., {Nelson}, D., {Pillepich}, A., {et~al.} 2019, arXiv
  e-prints, arXiv:1904.07238.
\newblock \doarXiv{1904.07238}

\bibitem[{{Wang} \& {Li}(2019)}]{2019MNRAS.483.1452W}
{Wang}, L., \& {Li}, C. 2019, \mnras, 483, 1452, \dodoi{10.1093/mnras/sty3204}

\bibitem[{{Weinberger} {et~al.}(2017){Weinberger}, {Springel}, {Hernquist},
  {Pillepich}, {Marinacci}, {Pakmor}, {Nelson}, {Genel}, {Vogelsberger},
  {Naiman}, \& {Torrey}}]{2017MNRAS.465.3291W}
{Weinberger}, R., {Springel}, V., {Hernquist}, L., {et~al.} 2017, \mnras, 465,
  3291, \dodoi{10.1093/mnras/stw2944}

\bibitem[{{Weinberger} {et~al.}(2018){Weinberger}, {Springel}, {Pakmor},
  {Nelson}, {Genel}, {Pillepich}, {Vogelsberger}, {Marinacci}, {Naiman},
  {Torrey}, \& {Hernquist}}]{2018MNRAS.479.4056W}
{Weinberger}, R., {Springel}, V., {Pakmor}, R., {et~al.} 2018, \mnras, 479,
  4056, \dodoi{10.1093/mnras/sty1733}

\bibitem[{{Weston} {et~al.}(2017){Weston}, {McIntosh}, {Brodwin}, {Mann},
  {Cooper}, {McConnell}, \& {Nielsen}}]{2017MNRAS.464.3882W}
{Weston}, M.~E., {McIntosh}, D.~H., {Brodwin}, M., {et~al.} 2017, \mnras, 464,
  3882, \dodoi{10.1093/mnras/stw2620}

\bibitem[{{White}(1994)}]{1994astro.ph.10043W}
{White}, S. D.~M. 1994, arXiv e-prints, astro.
\newblock \doarXiv{astro-ph/9410043}

\bibitem[{{Yang} {et~al.}(2019){Yang}, {Ge}, \& {Lu}}]{2019SCPMA..62l9511Y}
{Yang}, C., {Ge}, J., \& {Lu}, Y. 2019, Science China Physics, Mechanics, and
  Astronomy, 62, 129511, \dodoi{10.1007/s11433-019-9442-9}

\bibitem[{{Zakamska} {et~al.}(2019){Zakamska}, {Sun}, {Strauss}, {Alexandroff},
  {Brandt}, {Chiaberge}, {Greene}, {Hamann}, {Liu}, {Perrotta}, {Ross}, \&
  {Wylezalek}}]{2019MNRAS.489..497Z}
{Zakamska}, N.~L., {Sun}, A.-L., {Strauss}, M.~A., {et~al.} 2019, \mnras, 489,
  497, \dodoi{10.1093/mnras/stz2071}

\bibitem[{{Zel'Dovich}(1970)}]{1970A&A.....5...84Z}
{Zel'Dovich}, Y.~B. 1970, \aap, 500, 13

\bibitem[{{Zhao} {et~al.}(2019){Zhao}, {Ho}, {Zhao}, {Shangguan}, \&
  {Kim}}]{2019ApJ...877...52Z}
{Zhao}, D., {Ho}, L.~C., {Zhao}, Y., {Shangguan}, J., \& {Kim}, M. 2019, \apj,
  877, 52, \dodoi{10.3847/1538-4357/ab1921}

\end{thebibliography}



\end{document}